\documentclass{aa}  
\usepackage{graphicx}
\usepackage{xcolor}
\usepackage{txfonts}
\usepackage[english]{babel} 

\usepackage{adjustbox}

\usepackage{amsmath}
\usepackage{amsfonts}
\usepackage{amssymb}
\usepackage{wasysym}

\usepackage[flushleft]{threeparttable}

\usepackage{natbib}  

\begin{document}

   \title{Massive molecular gas reservoir around the central AGN in the CARLA J1103+3449 cluster at $z=1.44$. }
   \titlerunning{Massive molecular gas reservoir in cluster at $z=1.4$}

   \author{Vladan Markov
          \inst{1},
          Simona Mei\inst{2,3},
          Philippe Salom\'e\inst{1}, Francoise Combes\inst{1,4}, Daniel Stern\inst{3}, Audrey Galametz \inst{5}, Carlos De Breuck\inst{6}, Dominika Wylezalek\inst{6}, Stefania Amodeo\inst{1,7}, Elizabeth A. Cooke\inst{8},   Anthony H. Gonzalez\inst{9}, Nina A. Hatch\inst{10}, Ga\"el Noirot\inst{11},   Alessandro Rettura\inst{3}, Nick Seymour\inst{12}, Spencer A. Stanford\inst{13}, Jo\"el Vernet\inst{6}}
  \institute{LERMA, Observatoire de Paris, PSL Research University, CNRS, Sorbonne Universit\'e,  F-75014 Paris, France \email{vladan.markov@obspm.fr}
\and
Universit\'{e} de Paris, F-75013, Paris, France, LERMA, Observatoire de Paris, PSL Research University, CNRS, Sorbonne Universit\'e,  F-75014 Paris, France  \email{simona.mei@obspm.fr}
                        \and
             Jet Propulsion Laboratory, Cahill Center for Astronomy \& Astrophysics, California Institute of Technology, 4800 Oak Grove Drive, Pasadena, CA, USA
             \and
             Coll\'ege de France, 11 Place Marcelin Berthelot, 75231 Paris, France
             \and
             Department of Astronomy, University of Geneva, 1205, Versoix, Switzerland
             \and
           European Southern Observatory, Karl-Schwarzschildstrasse 2, 85748 Garching, Germany
           \and
                     Cornell Center for Astrophysics and Planetary Science, Cornell University, Ithaca, NY 14853, USA 
           \and
                      National Physical Laboratory, Hampton Road, Teddington, Middlesex, TW11 0LW, UK                  
           \and   
                      Department of Astronomy, University of Florida, Gainesville, FL 32611-2055, USA
                      \and
         School of Physics and Astronomy, University of Nottingham, University Park, Nottingham NG7 2RD, UK
                      \and
                      Department of Astronomy \& Physics, Saint Mary's University, 923 Robie Street, Halifax, NS B3H 3C3, Canada
                      \and
                                                                International Center for Radio Astronomy Research, Curtin University, GPO Box U1987, 6102 Perth, Australia
                      \and
                      Department of Physics, University of California, One Shields Avenue, Davis, CA 95616, USA
                      }

   \date{Received 16 June 2020; Accepted 2 July 2020 A\&A, 641, A22}
\authorrunning{Markov et al.}
 
  \abstract
  {Passive early-type galaxies dominate cluster cores at $z \lesssim1.5$. At higher redshift, cluster core galaxies are observed to have on-going star-formation, fuelled by cold molecular gas. 
   We measure the molecular gas reservoir of the central region around the radio-loud AGN in the cluster CARLA J1103+3449 at $z=1.44$ with NOEMA.
The AGN synchrotron emission dominates the continuum emission at 94.48 GHz, and we measure its flux at the AGN position and at the position of two radio jets. Combining our measurements with published results over the range 4.71~GHz-94.5~GHz, and assuming $\rm S_{synch} \propto \nu ^{-\alpha}$,  we obtain a flat spectral index $\alpha = 0.14 \pm 0.03$ for the AGN core emission, and a steeper index $\alpha = 1.43 \pm 0.04$ and $\alpha = 1.15 \pm 0.04$ at positions close to the western and eastern lobe, respectively.  The total spectral index is $\alpha = 0.92 \pm 0.02$ over the range 73.8~MHz-94.5~GHz. We detect two CO(2-1) emission lines, both blue-shifted with respect to the AGN. Their emission corresponds to two regions, $\sim17$~kpc southeast and $\sim14$~kpc southwest of the AGN, not associated with galaxies. In these two regions, we find a total massive molecular gas reservoir of $M^{\rm{tot}}_{\rm{gas}} = 3.9  \pm 0.4 \times 10^{10} M_{\astrosun}$, which dominates ($\gtrsim 60\%$) the central total molecular gas reservoir. These results can be explained by massive cool gas flows in the center of the cluster. The AGN early-type host is not yet quenched; its star formation rate is consistent with being on the main sequence of star-forming galaxies in the field (SFR$\approx 30-140 \ M_{\astrosun} {\rm  yr^{-1}}$), and the cluster core molecular gas reservoir is expected to feed the AGN and the host star formation before quiescence.  
The other cluster confirmed members show star formation rates at $\sim 2 \sigma$ below the field main sequence at similar redshifts and do not have molecular gas masses larger than galaxies of similar stellar mass in the field.}
 
   \keywords{galaxies: clusters: individual (CARLA J1103+3449) -- galaxies-evolution --
                galaxies-star formation -- galaxies-jets
                quenching}

   \maketitle
%

\section{Introduction} \label{intro}

At redshifts {\bf $z < 1.5$}  galaxy cluster cores are dominated by red, quenched, early-type galaxies, while blue, star-forming, late-type galaxies are mostly found in the field (e.g., \citealp{1980ApJ...236..351D}; \citealp{1998ApJ...504L..75B, 2004MNRAS.348.1355B}; \citealp{2005ApJ...623..721P}; \citealp{2009ApJ...690...42M}; \citealp{2011ApJ...732...94R}; \citealp{2012ApJ...745..106L, 2018arXiv181204624L}; \citealp{2015ApJ...800..107W}; \citealp{2018arXiv181204633T}).  At higher redshifts, results are somewhat conflicting, as it also becomes more difficult to define clusters of galaxies with measurements of their mass. Some results show that at $z > 1.5$ the star formation is already quenched in cluster cores (\citealp{2007MNRAS.377.1717K}; \citealp{2010A&A...524A..17S}; \citealp{2010ApJ...716.1503P}; \citealp{2012ApJ...756..114S}; \citealp{2012MNRAS.423.3652G}; \citealp{2012ApJ...753..164S}; \citealp{2012ApJ...756..115Z}; \citealp{2013ApJ...776....9G}; \citealp{2013ApJ...767...39M}; \citealp{2014ApJ...788...51N}; \citealp{2014ApJ...794..157M}; \citealp{2017ApJ...841L..21H}).
Other observations show a reversal of the star formation-density relation and ongoing star formation in cluster cores at $z > 1.5$, with a  much more varied galaxy population compared to clusters at lower redshifts (\citealp{2007A&A...468...33E}; \citealp{2008MNRAS.383.1058C}; \citealp{2010ApJ...719L.126T}; \citealp{2013ApJ...779..138B}; \citealp{2015MNRAS.447L..65S}; \citealp{2015ApJ...804..117M}; \citealp{2016ApJ...825...72A}; \citealp{2016ApJ...828...56W}; \citealp{2016ApJ...830...90N, 2018ApJ...859...38N}; \citealp{2018A&A...619A..49C}; \citealp{2018A&A...620A.198M}; \citealp{2018MNRAS.473.1977S};  \citealp{2018MNRAS.481.5630S}; \citealp{2019PASJ...71...40T}). A reversal of the star formation at $z\gtrsim1$ is also predicted from hydrodynamical and semi-analytical simulations (\citealp{2014ApJ...788..133T}; \citealp{2017ApJ...844L..23C}). 
Other cluster cores at $z \gtrsim1.5$ present equal percentages of quiescent and star-forming galaxies  (\citealp{2011A&A...527L..10F}; \citealp{2012MNRAS.423.2617T}; \citealp{2012ApJ...756..115Z}; \citealp{2012ApJ...754..141M}; \citealp{2016ApJ...830...90N}). 
A large presence of star-forming galaxies in cluster cores at $z\approx1.5-2$ indicates that most of the star formation quenching observed at lower redshift has not yet occurred, and that this is the key epoch of transformation of cluster galaxies from star-forming to passive.
At higher redshifts ($z\sim3-4$), proto-clusters show high star formation and star-burst activity (\citealp{2015ApJ...815L...8U}; \citealp{2018ApJ...862...96L}; \citealp{2018Natur.556..469M}; \citealp{2018ApJ...856...72O}; \citealp{2019ApJ...887..214K}; \citealp{2020MNRAS.495.3124H}; \citealp{2020arXiv200610753I}; \citealp{2020arXiv200313694L}; \citealp{2020ApJ...888...89T})

Galaxy star formation is fueled by cold and dense molecular gas (\citealp{2007ARA&A..45..565M}; \citealp{2012ARA&A..50..531K}; \citealp{2014PhR...539...49K}). Therefore, galaxies rich in cold molecular gas are mostly star-forming (bluish, mostly late-type spiral and irregular galaxies). Once the molecular gas is heated or stripped through different mechanisms, the star formation is quenched, and galaxies stop forming new, young, blue stars, which explode relatively fast due to their short life cycle. These galaxies will slowly become dominated by long-lived red stars, and galaxies will evolve into red, mostly elliptical, quenched galaxies (\citealp{1987gady.book.....B}; \citealp{1998ARA&A..36..189K}; \citealp{2012ARA&A..50..531K}).
There are several possible processes that can be responsible for star formation quenching, and each plays a different role in cold molecular gas removal, at different epochs and with different time scales (\citealp{2006PASP..118..517B}). Quenching depends on both  galaxy stellar mass and environment (\citealp{2004MNRAS.353..713K}; \citealp{2006MNRAS.373..469B}; \citealp{2010A&A...524A...2C}; \citealp{2010ApJ...721..193P, 2012ApJ...757....4P, 2014ApJ...790...95P}; \citealp{2013ApJS..206....3S}; \citealp{2015ApJ...805..121D, 2016ApJ...825..113D}; \citealp{2018arXiv180511475P}). More massive galaxy stellar populations are quenched at earlier epochs (\citealp{2005ApJ...621..673T}; \citealp{2013A&A...556A..55I}; \citealp{2013ApJ...777...18M}; \citealp{2013ApJ...772..113T}; \citealp{2015MNRAS.450.2749G}; \citealp{2016ApJ...832...79P}; \citealp{2016ApJ...817..118T}; \citealp{2017A&A...605A..29S}; \citealp{2017A&A...605A..70D}; \citealp{2018arXiv181206980M}; \citealp{2019A&A...621A..27F}).  Moreover, observations of  galaxies of the same stellar mass at $z<1.5$ show that the evolution from star-forming to quiescent is more rapid for cluster galaxies than for their field counterparts (\citealp{2011ApJ...732...12R}; \citealp{2012ApJ...746..188M}; \citealp{2012ApJ...750...93P}; \citealp{2013ApJ...770...58B}; \citealp{2013MNRAS.428..109S, 2014MNRAS.439.3189S}; \citealp{2013ApJ...772..118S}; \citealp{2013ApJS..206....3S}; \citealp{2014MNRAS.441..203D}; \citealp{2015MNRAS.450.2749G}; \citealp{2017MNRAS.464..876H}; \citealp{2017MNRAS.472.3512T}; \citealp{2018ApJ...866..136F}). This is due to the additional environmental mechanisms, such as tidal stripping (\citealp{1981ApJ...243...32F}; \citealp{1999MNRAS.304..465M}; \citealp{2018arXiv180506896C}), ram-pressure stripping (\citealp{1999MNRAS.308..947A}: \citealp{2008MNRAS.383..593M}; \citealp{2013MNRAS.429.1747M}; \citealp{2018MNRAS.476.4753J}), strangulation (\citealp{1980ApJ...237..692L}; \citealp{2000MNRAS.318..703B}; \citealp{2012ApJ...757....4P, 2015Natur.521..192P}; \citealp{2016A&A...590A.108M}) and galaxy merging in the first epochs of cluster formation (\citealp{2006ApJ...652..864H}; \citealp{2016MNRAS.463.3948D}). 

In the literature, the fraction of cold gas available for star formation is quantified as $f_{\rm{gas}}=M_{\rm{gas}}/(M_{*} + M_{\rm{gas}})$, or as a gas-to-stellar mass ratio $M_{\rm{gas}}/M_{*}$. These quantities depend on redshift, galaxy stellar mass, and environment. Observations have shown that for galaxies at a given stellar mass, the gas fraction and gas-to-stellar mass ratio increase with redshift (\citealp{2014ApJ...793...19S}; \citealp{2015ApJ...800...20G}; \citealp{2017ApJ...837..150S}; \citealp{2018ApJ...867...92S}; \citealp{2018ApJ...860..111D}; \citealp{2018ApJ...853..179T}). For galaxies at the same redshift, the gas fraction increases with decreasing stellar mass (\citealp{2013ApJ...768...74T, 2018ApJ...853..179T}; \citealp{2014ApJ...793...19S}; \citealp{0004-637X-842-1-55}). Finally,  at $ z<1.5$, for galaxies of the same mass and at the same redshift, cluster galaxies show  lower amounts of molecular gas and thus, lower gas fractions  (\citealp{2013A&A...557A.103J}; \citealp{2017ApJ...849...27R}; \citealp{0004-637X-842-1-55}; \citealp{2018arXiv180601826C}; \citealp{2018ApJ...856..118H}). Some works have shown that at higher redshifts ($z>2$), there is no difference in the gas fraction of cluster and field galaxies (\citealp{2016MNRAS.462..421H}; \citealp{2017A&A...608A..48D}).

In order to assess the molecular gas mass, we can estimate the mass of the most dominant interstellar molecule - $\rm{H_2}$, which is also the star formation fuel. However, this molecule is practically invisible to observations due to its lack of a permanent dipole moment and the fact that its rotational dipole transitions require high temperatures, $T > 100 \ \rm{K}$. In order to trace molecular hydrogen, rotational transitions of CO molecules are generally used for multiple reasons (\citealp{2012ARA&A..50..531K}, \citealp{2013ARA&A..51..105C}, \citealp{2013ARA&A..51..207B}). The CO molecule has a weak permanent dipole moment and it is easily excited even inside cold molecular clouds due to its low energy rotational transitions (\citealp{2012ARA&A..50..531K}; \citealp{2013ARA&A..51..207B}).  CO is also the second most abundant molecule after $\rm{H_2}$. CO rotational levels are excited by collisions with $\rm{H_2}$ molecules. Finally, CO rotational transitions lie in a relatively transparent millimeter window  (\citealp{2005ARA&A..43..677S}; \citealp{2012ARA&A..50..531K}). The main drawback with tracing molecular gas with CO line emission is that CO is a poor tracer of the so-called {\it CO-dark} molecular gas, which usually accounts for a significant fraction ($\sim 30\%$ to $\sim100\%$) of the total molecular gas mass in galaxies (\citealp{2005Sci...307.1292G}; \citealp{2010ApJ...716.1191W}; \citealp{2010ApJ...710..133A}; \citealp{2011A&A...536A..19P}; \citealp{2013A&A...554A.103P}; \citealp{2017ApJ...839...90B}; \citealp{2020arXiv200610154H}). In this paper, we focus on the molecular gas that can be detected by CO emission and molecular gas upper limits that can be inferred from the CO emission, with the caveat that this might not trace all the molecular gas in the galaxies that we study.

Few galaxy clusters are confirmed at $z \gtrapprox 1.5$. Current observations of the CO emission line in clusters at these epochs show that cluster galaxies still have cold gas to fuel their star formation. However, these results are not yet statistically significant, and some results point towards higher molecular gas content in cluster galaxies with respect to the field and others to lower (\citealp{2013A&A...558A..60C}; \citealp{2017ApJ...849...27R}; \citealp{2017ApJ...842L..21N}; \citealp{2018ApJ...856..118H}; \citealp{2018arXiv180509789C}; \citealp{2018arXiv180601826C}).
Molecular gas has also been detected in two protoclusters at $z \sim 2.5$ (\citealp{2015MNRAS.449L..68C}; \citealp{2016PhRvB..93q4104W}). Both protoclusters are dominated by star-forming (with a high starburst fraction), massive galaxies, with a substantial amount of molecular gas, and a small percentage of passive galaxies, which probably quenched after their accretion onto the cluster.

In this paper, we present IRAM (Institut de Radio Astronomie Millimetrique) NOEMA (NOrthern Extended Millimeter Array) observations of the core of a confirmed cluster from the CARLA (Clusters Around Radio-Loud AGNs;  \citealp{2013ApJ...769...79W}) survey at $z = 1.44$, CARLA J1103+3449 (\citealp{2018ApJ...859...38N}). CARLA J1103+3449 was selected as one of the highest CARLA IRAC color-selected overdensities ($ \sim 6.5 \sigma$, from \citealp{2014ApJ...786...17W}), and shows a $ \sim 3.5 \sigma$ overdensity of spectroscopically confirmed sources (our Fig. \ref{fig:cluster_noirot}, and Table~4 from  \citealp{2018ApJ...859...38N}). We find a large molecular gas reservoir south of the central AGN, consistent with gas inflows and outflows. We measure galaxy star formation rates and other properties for confirmed cluster members. We compare our results with similar observations in clusters and in the field. 

Our observations, data reduction and mapping are described in Sect. \ref{Data}, the results are given in  Sect. \ref{Results}, the discussion is in Sect. \ref{Discuss} and finally, the summary of our results is given in Sect. \ref{Summary}. Throughout this paper, we adopt a $\rm{\Lambda CDM}$ cosmology, with of $\Omega_{\rm M} = 0.3$, $\Omega_{\rm \Lambda} = 0.7$, $\Omega_{\rm k} = 0$ and $h = 0.7$, and assume a Chabrier initial mass function (IMF) (\citealp{2003PASP..115..763C}).

\section{Data}                      \label{Data}

\subsection{The CARLA survey} 

The CARLA survey \citep{2013ApJ...769...79W} is a substantial contribution to the field of high-redshift galaxy clusters at $z>1.5$. CARLA is a 408h Warm {\it Spitzer}/IRAC survey of galaxy overdensities around 420 radio-loud AGN (RLAGN). The AGNs were selected across the full sky, approximately half radio loud quasars (RLQs) and half radio galaxies  (HzRGs), and in the redshift range of $1.3 < z <3.2$. \cite{2013ApJ...769...79W} identified galaxies at $z>1.3$ around the AGNs in each field, using a color selection in the IRAC channel~1 ($\lambda =3.6\ \mu {\rm m}$;  IRAC1, hereafter) and channel~2 ($\lambda = 4.5\ \mu {\rm m}$; IRAC2, hereafter). They found that $92\%$ of the selected RLAGN reside in dense environments, with the majority ($55\%$) of them being overdense  at a $> 2\sigma$ level, and $10\%$ of them at a $> 5\sigma$ level, with respect to the field surface density of sources in the {\it Spitzer} UKIDSS Ultra Deep Survey (SpUDS; \citealp{2013ApJS..206...10G}), selected in the same way. 

A {\it Hubble Space Telescope} Wide Field Camera 3 ({\it HST}/WFC3) follow-up of the twenty highest CARLA {\it Spitzer} overdensities (consisting of 10 HzRGs and 10 RLQs) spectroscopically confirmed sixteen of these  at $1.4 < z < 2.8$,  and also discovered and spectroscopically confirmed seven serendipitous structures at $0.9 < z < 2.1$  \citep{2018ApJ...859...38N}. The structure members were confirmed as line-emitters (in H${\alpha}$, H${\beta}$, [\ion{O}{II}], and/or [\ion{O}{III}], depending on the redshift) and have star formation estimates from the line fluxes \citep{2018ApJ...859...38N}. The star-formation of galaxies with stellar mass $\gtrsim 10^{10} M_{\astrosun}$ is below the star-forming {\it main sequence} (MS) of field galaxies at similar redshift, and star-forming galaxies are mostly found within the central regions \citep{ 2018ApJ...859...38N}.  This program also provided WFC3 imaging in the F140W filter (WFC3/F140W) from which we obtained point spread function (PSF) matched photometric catalogs (Amodeo et al., in preparation), and galaxy visual morphologies (Mei et al., in preparation).

 From their IRAC luminosity function, \citet{2014ApJ...786...17W} showed that CARLA overdensity galaxies have probably quenched faster and earlier compared to field galaxies. Some of the CARLA northern overdensities were also observed in either deep {\it z}-band or deep {\it i}-band, with GMOS at the Gemini telescope (hereafter Gemini/GMOS), ISAAC at the European Southern Observatory Very Large Telescope (VLT/ISAAC) and ACAM at the WHT (William Herschel Telescope) telescope (WHT/ACAM). This permitted us to estimate their galaxy star formation rate histories, and we deduced that, on average,  the star formation of galaxies in these targets had been rapidly quenched, producing the observed colors and luminosities (\citealp{2015MNRAS.452.2318C}).

\subsection{Optical and near-infrared multi-wavelength observations of CARLA J1103+3449 } \label{carla}

As a target of the {\it Spitzer} CARLA survey, CARLA J1103+3449 was observed with {\it Spitzer} IRAC1 and IRAC2 (Cycle 7 and 8 snapshot program; P.I.: D. Stern), for a total exposure of 800s and 2000s, respectively.  The IRAC cameras have $256 \times 256$ InSb detector arrays with a pixel size of 1.22 arcsec and a field of view of $5.2 \times 5.2$ arcmin. \citet{2013ApJ...769...79W} performed the data calibration and mapping with the \texttt{MOPEX} package (\citealp{2005ASPC..347...81M}) and detected sources with SExtractor (\citealp{1996A&AS..117..393B}), using the IRAC-optimized SExtractor parameters from the work of \cite{2005ApJS..161...41L}. The final {\it Spitzer} IRAC1 and IRAC2 mosaic has a pixel size of 0.61~\rm{arcsec}, after taking into account dithering and sub-pixelation.

The {\it HST}/WFC3 imaging and grism spectroscopy were obtained with the dedicated {\it HST} follow-up program (Program ID: 13740; P.I.: D. Stern). We obtained F140W imaging (with a field of view of $2 \times 2.3 \ \rm{arcmin^2}$ at a resolution of $0.06 \ \rm{arsec \ pix^{-1}}$, obtained after taking into account dithering), and G141 grism spectroscopy (with a thoughtput > 10\% in the wavelength range of $1.08 \ \mu {\rm m }< \lambda < 1.70 \ \mu {\rm m }$ and spectral resolution $R = \lambda/\Delta \lambda = 130$).  This grism was chosen in order to permit the identification of strong emission lines at our target redshift, such as  H$\alpha$, H$\beta$, [\ion{O}{II}] and [\ion{O}{III}]. \citet{2016ApJ...830...90N, 2018ApJ...859...38N} performed the data reduction using the  \texttt{aXe} (\citealp{2009PASP..121...59K}) pipeline,  by combining the individual exposures, and removing cosmic ray and sky signal.  \citet{2018ApJ...859...38N}  performed the source detection with SExtractor (\citealp{1996A&AS..117..393B})  and extracted two-dimensional spectra for each field, based on the positions and sizes of the sources. The redshifts and emission line fluxes were determined using the python version of \texttt{mpfit} and are published in  \cite{2018ApJ...859...38N}. 

CARLA J1103+3449 was followed-up with {\it i}-band imaging with WHT/ACAM  (PI: N. Hatch; \citealp{2015MNRAS.452.2318C}), and we obtained a PSF--matched photometric catalog in the WHT/ACAM {\it i}-band, WFC3/F140W (detection image), IRAC1 and IRAC2. The {\it i}-band, WFC3/F140W, and IRAC1 filters correspond to the {\it UVJ} rest-frame bandpasses at the redshift of CARLA J1103+3449.

More details on the {\it Spitzer} IRAC, {\it HST}/WFC3 and WHT/ACAM   observations, data reduction and results can be found in \cite{2013ApJ...769...79W, 2014ApJ...786...17W}, \cite{2016ApJ...830...90N, 2018ApJ...859...38N}, and \citet{2015MNRAS.452.2318C}, respectively.

From a morphological (from the {\it HST}/F140W images) and photometric analysis of the central sources  (Amodeo et al., in preparation;  Mei et al., in preparation), the host galaxy of the AGN is an elliptical galaxy. The spiral galaxy close to the AGN is a spectroscopically confirmed member \citep{2018ApJ...859...38N}, but is not detected as an independent galaxy in the IRAC images because of their poor spatial resolution. The bright central source south of the AGN is a star, with a spectral energy distribution consistent with a black body and not consistent with an early-type galaxy (ETG) spectrum. 

\subsection{Keck AGN spectrum observations}

The redshift for the radio source B2~1100+35, associated with
WISE~J110326.19+344947.2 at the center of CARLA~J1103+3449, was
first reported in \cite{1997MNRAS.291..593E} as $z =
1.44$, but with no spectrum presented.  With no spectrum available
from the Sloan Digital Sky Survey of the faint, red ($g = 23.9$~mag,
$i = 21.4$~mag) optical counterpart to the radio source, we observed
B2~1100+35 with the dual-beam Low Resolution Imaging Spectrometer
(LRIS; \citealp{1995PASP..107..375O}) at Keck Observatory on UT
2019 March 10.  The night suffered strongly from variable, often
thick cloud cover.

\begin{figure}[h]
\centering
\includegraphics[width=\hsize]{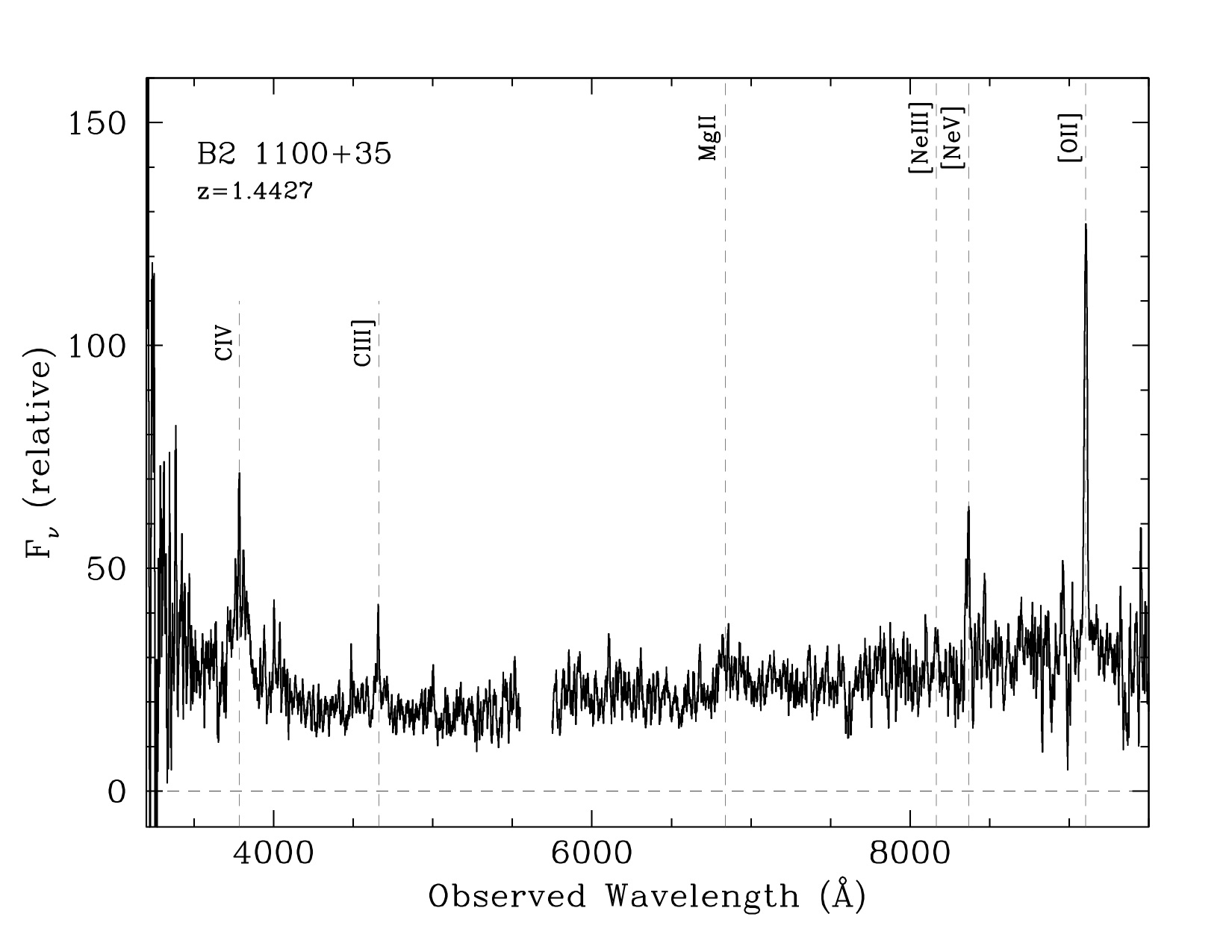}
\caption{\textit {The Keck/LRIS spectrum of B2~1100+35, the radio galaxy at the center
of CARLA~J1103+3449.  Since the night was not photometric, the
y-axis only provides relative flux calibration.}}
\label{fig:keck}
\end{figure}

The data were obtained through the 1\farcs0 slit with the 5600~\AA\
dichroic.  The blue arm of the spectrograph used the 600~$\ell\,
{\rm mm}^{-1}$ grism ($\lambda_{\rm blaze} = 400$~\AA; resolving
power $R \equiv \lambda / \Delta \lambda \sim 1600$ for objects
filling the slit), while the red arm used the 4000~$\ell\, {\rm
mm}^{-1}$ grating ($\lambda_{\rm blaze} = 8500~$\AA; $R \sim 1300$).
Three 600~s exposures were attempted, though ultimately only one
proved useful.  We processed the spectrum using standard techniques,
and flux calibrated the spectrum using observations of the standard
stars Hilter~600 and HZ44 from \citet{1990ApJ...358..344M} obtained the same night with the same instrument configuration.
Fig.~\ref{fig:keck} presents the processed spectrum.  Multiple redshifted
emission lines are detected, including broadened \ion{C}{IV}~$\lambda
1549$~\AA, narrow \ion{C}{III}]~$\lambda 1909$~\AA, narrow
[\ion{Ne}{V}]~$\lambda 3426$~\AA, and strong, narrow [\ion{O}{II}]~$\lambda
3727$~\AA.  Based on the latter feature, we report a redshift of
$z = 1.4427 \pm 0.0005$, where the uncertainty reflects both statistical
uncertainties in the line fitting, as well as an estimate of
systematic uncertainties in the wavelength calibration, and a
comparison with other well-detected emission lines in this source.
This measurement is consistent with the \citet{2018ApJ...859...38N} AGN redshift measurement of  $z = 1.444 \pm 0.006$, from the {\it HST}/WFC3 grism observations (see above).

\subsection{IRAM observations}

For this work, we observe CARLA J1103+3449 with the IRAM/NOEMA (P.I.: A. Galametz, S. Mei), with eight antennas over a five day period (28--30 July, 3--4 August 2017), for a total exposure time of $\sim $29 h (including overheads). The weather conditions were within the  average ($\rm{PWV}$ $\sim 10-20$ mm). The average system temperature was $T_{\rm{sys}}$~$\sim 100-200$ K, and reached maximum values of 300~K. The sources used as RF (receiver bandpass) calibrator, the flux calibrator and amplitude/phase calibrators were the 3C84 radio galaxy, the LKHA101 radio star, and the 1128+385 quasar, except on the $30^{\rm{th}}$ of July, when we used the quasars 3C273, 1128+385 (measured on the $28^{\rm{th}}$ of July) and 1156+295.

\begin{figure}[h]
\centering
\includegraphics[width=\hsize]{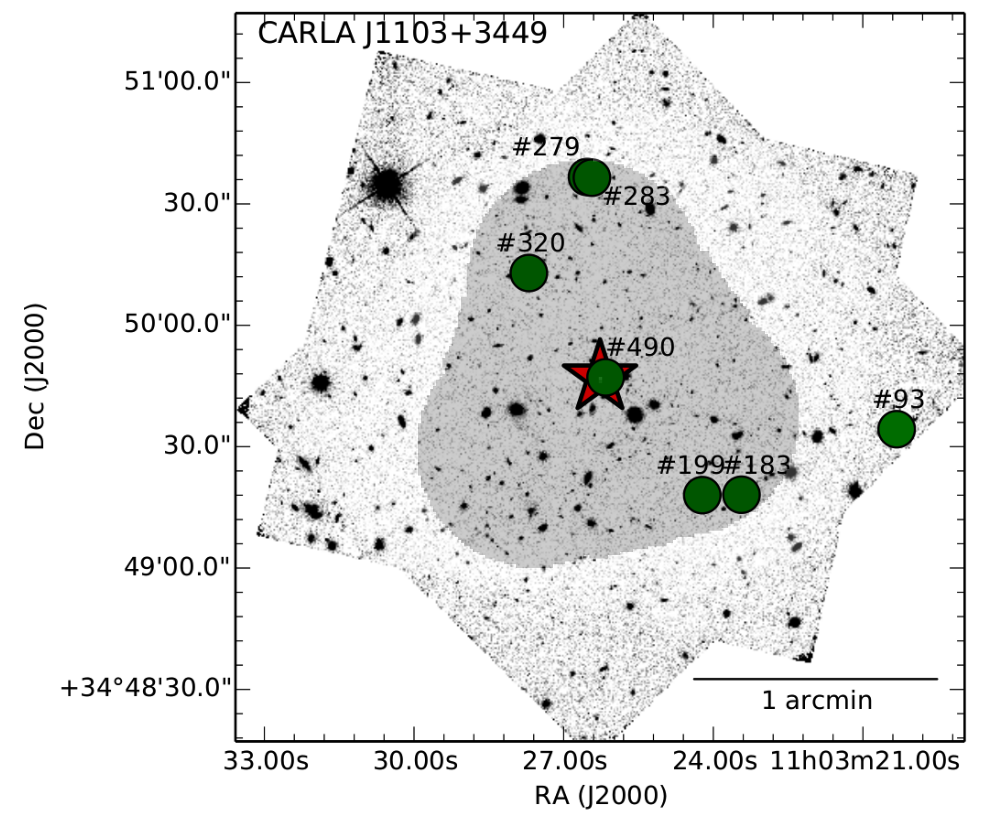}
\caption{\textit {The distribution of spectroscopically confirmed members of  CARLA J1103+3449 (green circles) with the central AGN (red star) \citep{2018ApJ...859...38N}. The background images are the two orientation HST/WFC3 F140W frames. The shaded area indicates the NOEMA mosaic map area.}}
\label{fig:cluster_noirot}
\end{figure}

We target the CO(2-1) emission line at the rest-frame frequency $\nu_{\rm{rest}} = 230.538\ \rm{GHz}$, which is redshifted to $\nu_{\rm{obs}} = 94.48\ \rm{GHz}$ at $z \sim 1.44$ (the mean confirmed cluster member redshift from \citealp{2018ApJ...859...38N}), observed in NOEMA's $3$mm band. We cover our target with three pointings to map the AGN and the central cluster region. The pointings were positioned so that we could cover as many IRAC color-selected members as possible ($\sim40$) along with the 7 (out of 8) {\it HST}/WFC3 spectroscopically confirmed members (green circles and a red star, for the AGN, in Fig. \ref{fig:cluster_noirot}, based on \citealp{2018ApJ...859...38N}). We chose the antenna configuration C to be able to separate cluster members in the cluster core. The beam size is  $4.14 \times 3.46$ arcsec$^2$, the $PA = -171.01^{\circ}$ and the velocity resolution is $50 \ \rm{km \ s^{-1}}$ (smoothed to $100 \ \rm{km \ s^{-1}}$; see below).

We perform the entire NOEMA data calibration by running the pipeline in the \texttt{clic} package of the IRAM/GILDAS software\footnote{http://www.iram.fr/IRAMFR/GILDAS/}. We flag additional data, modify antenna positions and calibrate again the flux. While the pointing and the focus were excellent, the amplitude and phase were of average quality because of the weather conditions. 

\begin{figure*}[h]
\centering
\includegraphics[width=0.49\textwidth]{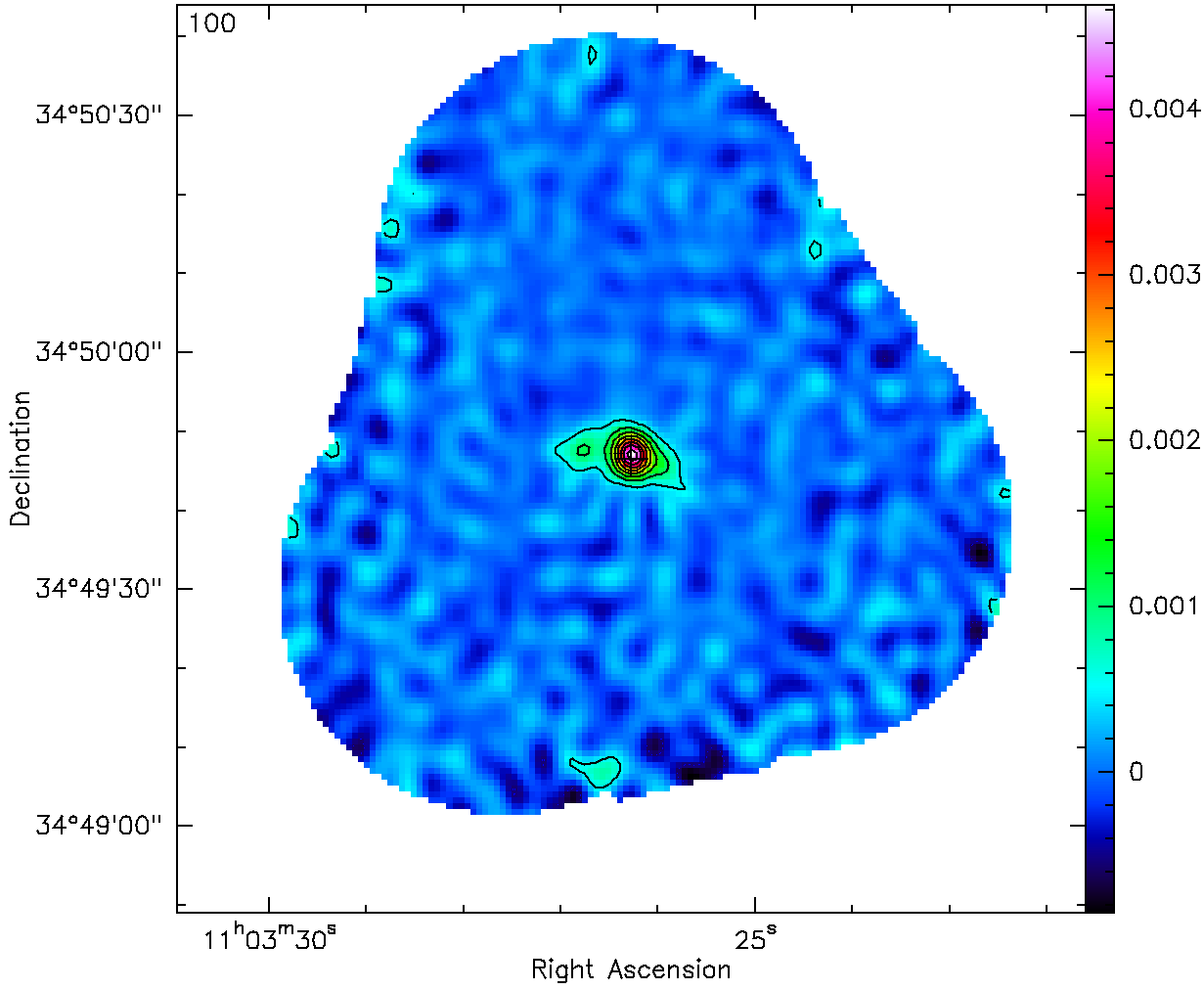}
\includegraphics[width=0.50\textwidth]{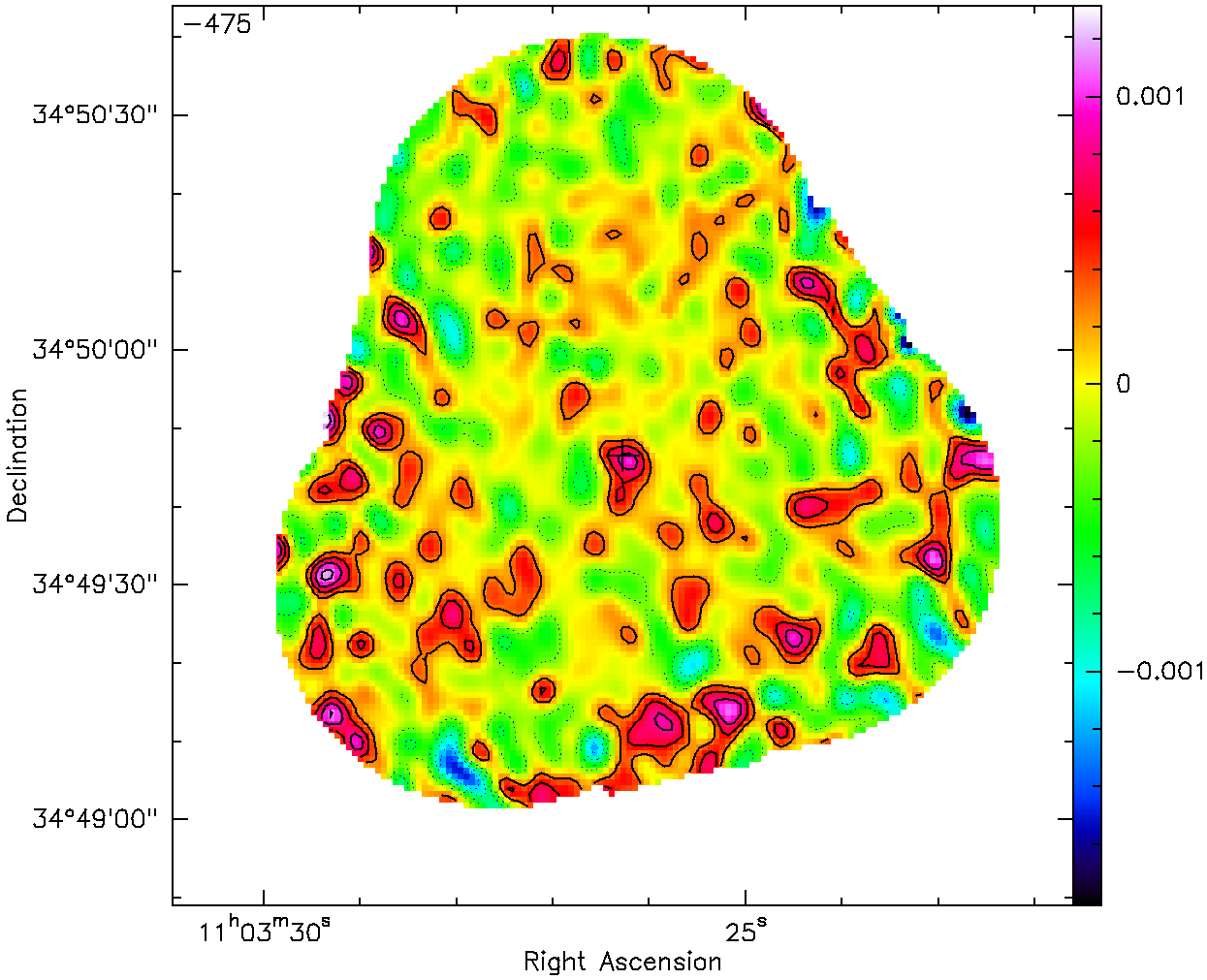}
\caption{\textit {Left : the continuum emission map of the CARLA J1103+3449 cluster at 94.48 GHz. The map was obtained by averaging the flux over a velocity range of $2450 \ km \ s^{-1}$, outside of the emission lines. The rms noise level is $\sigma \sim 0.2 \ \rm{mJy \ beam^{-1}}$ in the central pointing intersection region, and the  contours show the $3\sigma$, $6\sigma$, $9\sigma$, etc.  levels up to $24\sigma$. Right: the continuum subtracted CO(2-1) line emission mosaic map. The color wedge of the intensity maps is in $\rm{Jy \ beam^{-1} }$. The map was obtained by averaging the flux over a velocity range of $1200 \ km \ s^{-1}$, and has an average rms noise level of $\sigma \sim 0.2 \ \rm{mJy \ beam^{-1}}$ in the central pointing intersection region. The continuous lines show positive $\sigma$ contours and the dotted lines show negative $\sigma$ contours. The contours show the $1\sigma$, $2\sigma$ and $3\sigma$ levels. The cross marks the phase center of the mosaic.  The noise approximately doubles towards the map edges because of the primary beam correction. Both maps show an extended source in the cluster center.  }}
\label{fig:continuum}
\end{figure*}

With the reduced data, we create a CO(2-1) continuum emission mosaic map, by using the \texttt{mapping} package of the GILDAS software (Fig. \ref{fig:continuum} left). 
The map was obtained by averaging the flux over a velocity range of $2450 {\rm \ km \ s^{-1}}$, excluding emission lines, with a background rms noise level of $\sigma \sim 0.2 \ \rm{mJy \ beam^{-1}}$. 
Then, we subtract the continuum from the CO(2-1) emission in the uv-data set in order to obtain a clean, continuum-subtracted CO(2-1) map. 

We calculate the root mean square (rms) noise level in the three pointing intersection region of the CO(2-1) map (see Fig.~\ref{fig:cluster_noirot}). In this region, the original velocity resolution is $50\ \rm{ km \ s^{-1}}$ and the rms noise level is $\sigma \sim 0.8 \times \rm{mJy \ beam^{-1}}$. 
In order to improve the signal-to-noise ratio (SNR), we smooth the CO(2-1) map to a final velocity resolution of $100\ \rm{km \ s^{-1}}$ by averaging two consecutive channels, and obtain a rms noise level of $\sigma \sim 0.5 \times \rm{mJy \ beam^{-1}}$ after smoothing. We create the CO(2-1) intensity map by averaging the flux over a velocity range of $1200 \ \rm{ km \ s^{-1}}$ with the background rms noise level of $\sigma \sim 0.2 \ \rm{mJy \ beam^{-1}}$, and apply a primary beam correction (Fig. \ref{fig:continuum}, right). On the mosaic edges, the rms noise level approximately doubles.

\begin{figure*}[h]
\centering
\includegraphics[width=0.6\textwidth]{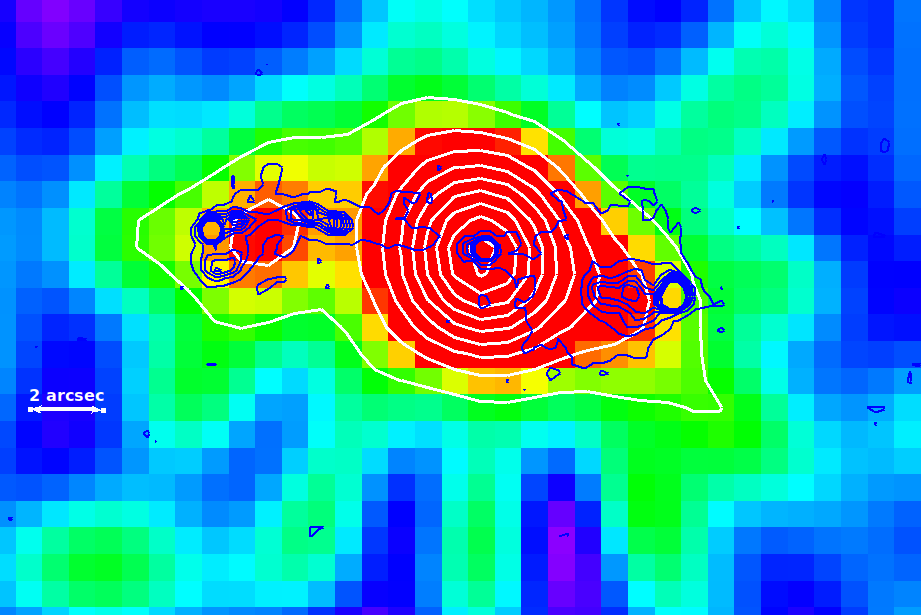}
\caption{\textit {A zoom-in on the continuum emission map of the extended source in the cluster center. The continuum emission contours (white) run as $3\sigma$, $6\sigma$, $9\sigma$, etc. up to $24\sigma$. The radio emission contours at $4.71$ GHz from the work of \cite{1999MNRAS.303..616B} are overlaid on the image (in blue). The brightest continuum emission peak and one of the radio peaks are both centered on the AGN. The continuum emissions visually corresponds to the position of the radio jets, suggesting  the same or a connected physical origin. North is up and East is to the left. }}
\label{fig:contem}
\end{figure*}  
\section{Results} \label{Results}

\subsection{THE AGN CONTINUUM EMISSION}

On the continuum emission map at the observed frequency of $\nu_{\rm{obs}} = 94.48\ \rm{GHz}$, we detect an extended source in the cluster central region, with the brightest peak at the position of the AGN (>26$\sigma$, Fig. \ref{fig:contem}, white contours). 
Comparing the NOEMA continuum emission with radio observations at $4.71$ GHz from \cite{1999MNRAS.303..616B} (Fig. \ref{fig:contem}, blue contours), the NOEMA continuum emission visually corresponds to the radio jets. Both the NOEMA extended continuum emission peak, and the central radio emission, correspond to the AGN position. We also detect significant (>6$\sigma$) continuum emission at the position of the tip of the eastern radio lobe (Fig. \ref{fig:contem}). 

The  position and the scale of this continuum emission detection follow the emission from the radio lobes, with the brighter and the fainter continuum components corresponding to the western and eastern radio lobe, respectively. This is consistent with the same or a connected physical origin of the two emissions. In fact, the AGN synchrotron emission dominates at both our NOEMA observed frequency  (rest-frame $\nu \sim 230\ \rm{GHz}$/$\lambda \sim 1.3\ \rm{ mm}$), and in the radio observation frequency range (\citealp{1990A&ARv...2..125B}; \citealp{1998ApJ...503L.109H};   \citealp{2008A&A...485...33H}; \citealp{2017ApJ...845...50N}; \citealp{2019MNRAS.484.4239R}). At both frequency ranges, the signal corresponds to the non-thermal synchrotron radiation emitted by the relativistic charged particles from the AGN jets (\citealp{1995ApJ...449L..19G, 1997ApJ...482L..33G}; \citealp{1997ApJ...476..649M}; \citealp{2000ApJ...528L..85A}; \citealp{2011ApJ...737...42P};  \citealp{2018ApJ...860..121F}).

We measure the continuum within a region the size of the NOEMA beam centered on the AGN and the two lobes. We obtain $S^{\rm{AGN}}_{\rm{cont}} = 4.6 \pm 0.2 \ \rm{mJy}$, $S^{\rm{east\_lobe}}_{\rm{cont}} = 1.1 \pm 0.2 \ \rm{mJy}$, and $S^{\rm{west\_lobe}}_{\rm{cont}} = 0.8 \pm 0.2 \ \rm{mJy}$. For the eastern lobe, we centered our measurement on the peak of the emission in our observations, while for the western lobe, where we don't have a clear peak, we centered on the radio peak.

\begin{table*}[!h]
\center
\caption{Continuum flux measurements}
\begin{tabular}{c c c c c c c c}
\hline \hline
\\
Component & RA (J2000) & DEC (J2000) & $F_{4.71 \ \rm GHz}$ &$F_{8.21 \rm \ GHz}$ & $F_{94.5 \ \rm GHz}$ & $\alpha^{\rm{B99}}$ & $\alpha$  \\
&(h:m:s)&(d:m:s)&  (mJy) &(mJy) & (mJy) & &  \\
\\
\hline total &...&...& 96.6 & 55.7 & $6.5\pm 0.3$ &...& $0.94 \pm 0.01$ \\
\hline core &  11:03:26.26 & +34:49:47.2 & 6.8 & 6.9 & $4.6 \pm 0.2$ & -0.04 & $0.14 \pm 0.03$ \\
\hline west & 11:03:25.83 & +34:49:45.9 & 57.0 & 29.1 & $0.8 \pm 0.2$ & 1.21& $1.43 \pm 0.04$\\
\hline east+jet &  11:03:26.77 & +34:49:47.7 & 32.8 & 19.7 & $1.1 \pm 0.2$ &...& $1.15 \pm 0.04$ \\
\hline east &  11:03:26:89 & +34:49:47.8 & 21.7 & 12.7 &...& 0.96 &...\\
\hline jet & 11:03:26.64 &+34:49:48.2 &11.1 & 7.0 &...& 0.82 &...\\
\hline
\end{tabular}
\label{tab:A}
\tablefoot{Continuum flux measurements from our work at 94.5 GHz, and those from \cite{1999MNRAS.303..616B} at 4.71~GHz and 8.21~GHz. $\alpha$ is the spectral index that we measure over this wavelength range, and $\alpha^{\rm{B99}}$ are the spectral indexes from \cite{1999MNRAS.303..616B}.}
\end{table*}

Comparing the continuum emission from this work and the radio emission from \cite{1999MNRAS.303..616B} (white and blue contours in Fig. \ref{fig:contem}, respectively), we note that our brighter continuum component ($\sim 83\%$ of the total continuum emission flux) roughly corresponds to the radio emission of the core and the western lobe  ($\sim 65\%$ at $\nu = 8210\ \rm{MHz}$ and $66\%$ at $\nu = 4710\ \rm{MHz}$ of the total flux), and also includes the fainter part of the eastern jet. The peak of the continuum emission at 94.48 $\rm{GHz}$ is centered on the AGN ($\sim 71\%$ of the total continuum emission flux), while most of the radio emission from \cite{1999MNRAS.303..616B} is from the western lobe ($\sim 52\%$ at $\nu = 8210\ \rm{MHz}$ and $59\%$ at $\nu = 4710\ \rm{MHz}$ of the total flux).
Our fainter continuum component ($\sim 17\%$ of the total flux) corresponds to the radio emission of the eastern lobe and a brighter part of the eastern jet ($\sim 35\%$ at $\nu = 8210\ \rm{MHz}$ and $34\%$ at $\nu = 4710\ \rm{MHz}$ of the total flux).

Table \ref{tab:A} shows our continuum flux measurements at 94.48~GHz, and those from \cite{1999MNRAS.303..616B} at 4.71~GHz and 8.21~GHz.  The total flux in the table is the sum of the three components, the core and the two lobes. The total flux measured in an area with the signal exceeding 3~$\sigma$ of the background is $S^{\rm{3\sigma}}_{\rm{cont}} = 8.2 \pm 0.2 \ \rm{mJy}$.  We model the AGN and lobe spectral energy distribution (SED) as a power law ($S_{\rm{synch}} \propto \nu^{-\alpha}$), and obtain the spectral index $\alpha$ for the different components from a linear fit in logarithmic scale, using all three frequencies. The uncertainty on $\alpha$ is the statistical uncertainty from the linear fit. The systematic uncertainty on $\alpha^{\rm{B99}}$ of $0.07$ from  \cite{1999MNRAS.303..616B} is much larger than the statistical uncertainty and is calculated by assuming $3\%$ uncertainties in the absolute calibration at each \cite{1999MNRAS.303..616B} frequency. Our indexes are consistent (1-1.5~$\sigma$) with those from \cite{1999MNRAS.303..616B} (also shown in Table \ref{tab:A}). The lobes present a steep SED,  consistent with the jets' optically thin synchrotron emission (\citealp{1999MNRAS.303..616B}; \citealp{10.1093/mnras/stt531}; \citealp{2017ApJ...845...50N}; \citealp{2019MNRAS.484.4239R}; \citealp{2019MNRAS.488.1917G}), while the AGN core SED is flatter, which is consistent with optically thicker (self-absorbed) synchrotron emission (\citealp{1999MNRAS.303..616B}; \citealp{2019MNRAS.484.4239R}; \citealp{2019MNRAS.488.1917G}).

\begin{figure}[h]
\centering
\includegraphics[width=0.7\hsize]{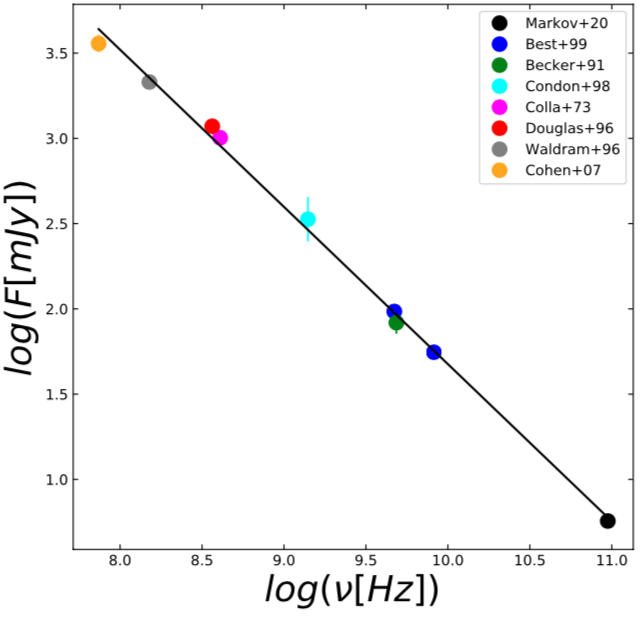}
\caption{\textit {Total  {\rm SED} plot of the AGN in the radio and mm wavebands from our work (black),  \cite{1999MNRAS.303..616B} (blue), \cite{1991ApJS...75....1B} (green), \cite{1998AJ....115.1693C} (cyan),
\cite{1973A&AS...11..291C} (magenta), \cite{1996AJ....111.1945D} (red), \cite{1996MNRAS.282..779W} (grey) and \cite{2007AJ....134.1245C} (orange). Over this range of frequencies, we obtain an AGN spectral index of $\alpha = 0.92 \pm 0.02$}}
\label{fig:synchro}
\end{figure}

 In Fig.~\ref{fig:synchro} we show the total AGN spectral energy distribution (SED) at radio and mm wavelengths from our work (total AGN emission) and the literature. Over this larger range of frequencies, we obtain $\alpha = 0.92 \pm 0.02$, consistent with the optically thin synchrotron emission of AGN jets that dominate the total continuum emission. The SED does not show a flattening or steepening neither at high-frequency ($\nu > 10$ GHz), in agreement with previous results (\citealp{2006MNRAS.371..852K}; \citealp{2011ApJ...734L..25E}; \citealp{2019A&A...621A..27F}).
 
 Our results are within the range of spectral indexes found in previous work. The typical spectral index of optically thin synchrotron emission (which corresponds to jets) is in the range of $0.5 \lesssim \alpha \lesssim 1.5$ in the local Universe (\citealp{10.1093/mnras/stt531}; \citealp{2017ApJ...845...50N}; \citealp{2019MNRAS.484.4239R}; \citealp{2019MNRAS.488.1917G}) and  $1 \lesssim \alpha \lesssim  2$  for galaxies at $z>2$, with higher values being rarer (\citealp{1997ApJS..109....1C};  \citealp{1999MNRAS.303..616B}; \citealp{2019A&A...621A..27F}).
For the optically thick emission (which corresponds to the core), $-0.5 \lesssim \alpha \lesssim0.5$ is found in the local Universe (\citealp{2019MNRAS.484.4239R}; \citealp{2019MNRAS.488.1917G}) and $-1 \lesssim \alpha \lesssim 1$  is found at $z>2$, with most of the measurements being $\alpha > 0.5$  (\citealp{1997ApJS..109....1C}; \citealp{1997MNRAS.289..525A}; \citealp{1999MNRAS.303..616B}; \citealp{2019A&A...621A..27F}).

\subsection{THE MOLECULAR GAS CONTENT AROUND THE AGN}

\paragraph{System velocity and FWHM.} In order to estimate the system velocity and the velocity Full Width Half Maximum (FWHM) from the CO(2-1) line emission, we use the \texttt{class} package from the GILDAS software. We extract the CO(2-1) line profile from a polygon enclosing all $> 1\sigma$ CO(2-1) emission in the central region of the cluster.

\begin{figure*}[h]
\centering
\includegraphics[width=0.40\textwidth]{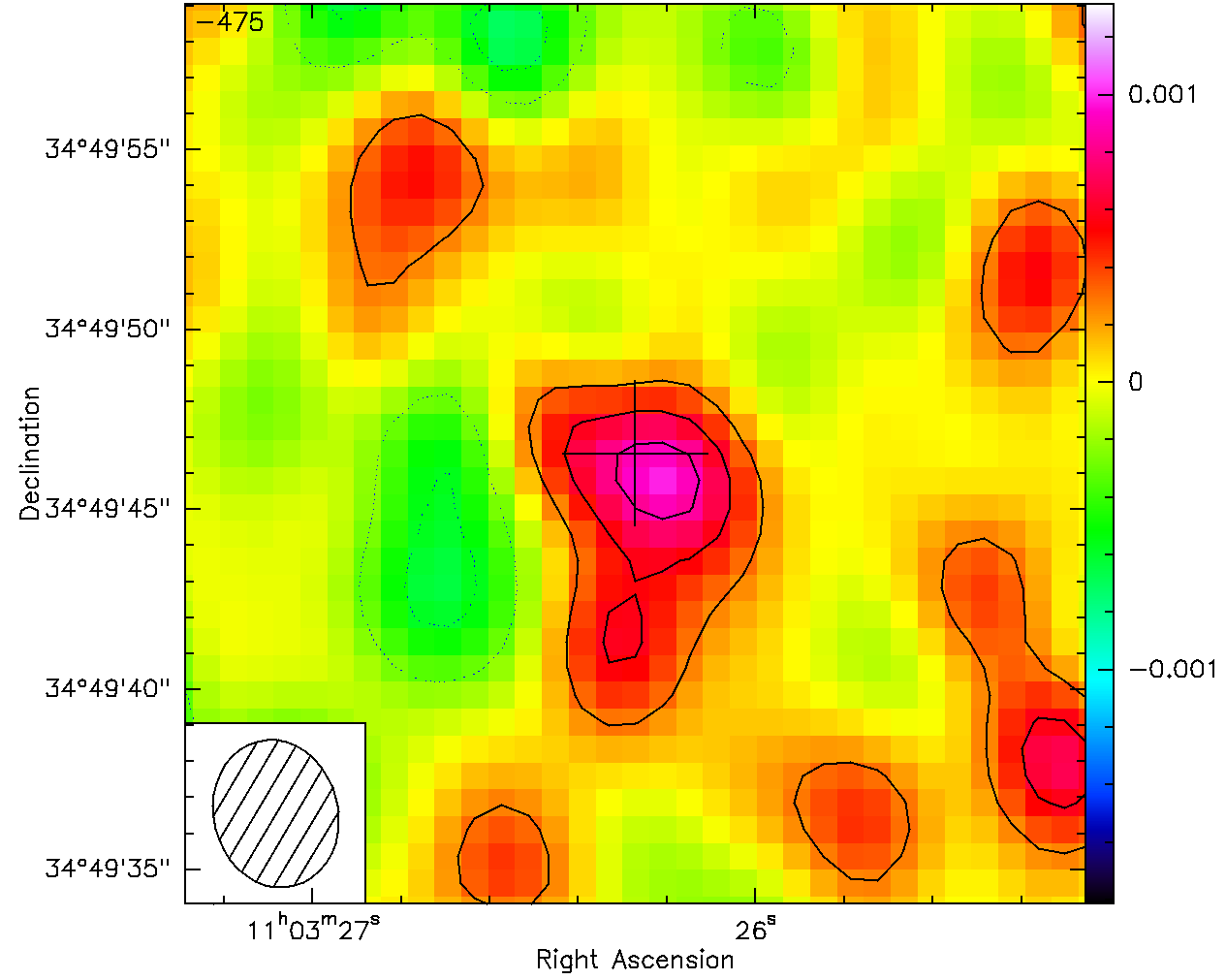}
\includegraphics[width=0.59\textwidth]{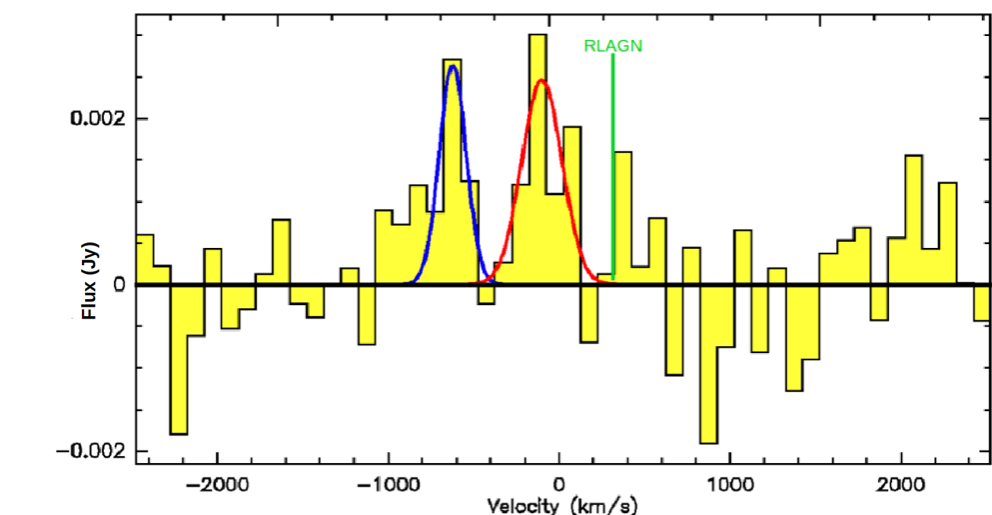}
\caption{\textit { Left: a zoom-in on the CO(2-1) line emission, continuum-subtracted mosaic map of the extended source in the cluster center. The black cross marks the center of our observations. The beam size ($4.14 \times 3.45$ arcsec$^2$) is plotted at the lower left. The color scale of the intensity map is in $\rm{Jy \ beam^{-1}}$. The rms noise level is $\sigma \sim 0.2 \ \rm{mJy \ beam^{-1}}$ in the central pointing intersection region. The contours correspond to the $1\sigma$, $2\sigma$ and $3\sigma$ levels. Right: the CO(2-1) line emission integrated spectrum. The two Gaussians fits correspond  to system velocities of $V{\rm{sys}} = -623.0\ \rm{ km\ s^{-1}}$ (blue), and $V{\rm{sys}} = -115.5\ \rm{ km \ s^{-1}}$ (red). The AGN spectroscopic redshift corresponds to a velocity of $v = 331.6 \  \rm{ km\ s^{-1}}$ (vertical green line). 
}}
\label{fig:mosaic}
\end{figure*}

\begin{figure*}[h]
\centering
\includegraphics[width=0.7\hsize]{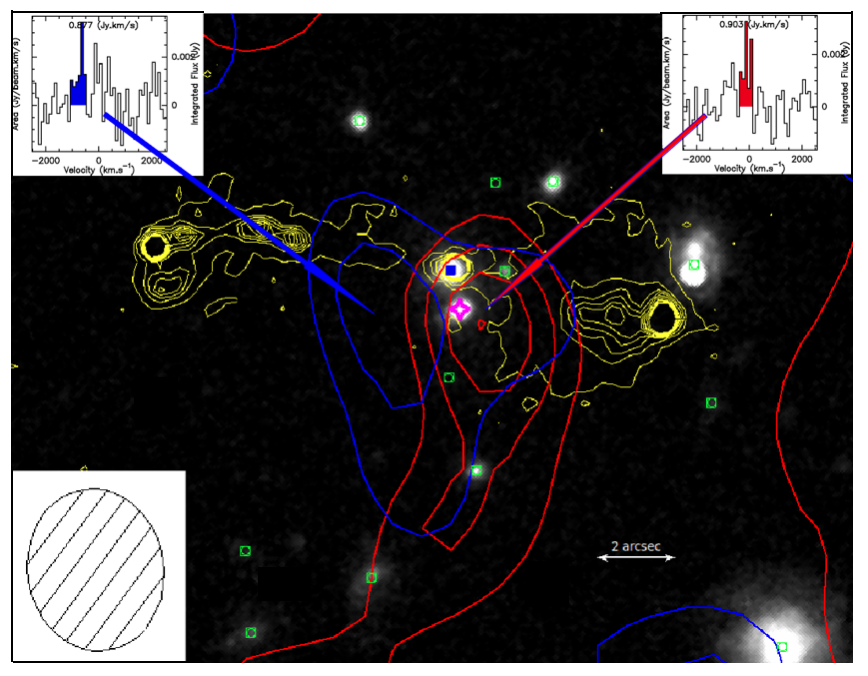}
\caption{\textit {The {\it HST}/WFC3 F140W image of the central region of the cluster, with contours of CO(2-1) emission of the eastern and western emission peaks (shown as blue and red contours, respectively), and the radio emission at $4.71$ GHz (yellow) from the work of \cite{1999MNRAS.303..616B}.  The central peak of the radio emission corresponds to the position of the AGN. The two radio lobes are asymmetrical, with the one to the east (left of the AGN) being more compact and the one to the west more diffuse. The central blue square and the pink star with four spikes show the position of the AGN and of the star, respectively. The green squares are the positions of IRAC color-selected galaxies in the cluster central region. The contours are derived by integrating the CO(2-1) emission across the velocities marked by their corresponding color on the CO(2-1) emission line spectra.  The spectra are shown on the top left and right insets (see Appendix \ref{peaks} for more details). The contour levels of the eastern and western emission peaks are $1-2 \ \sigma$ and $1-4 \ \sigma$, respectively.  The eastern and western emission peaks are south of the AGN, and do not correspond to any galaxy detected on the {\it HST} or {\it Spitzer} images. North is up and East is to the left. The beam scale is shown on the bottom left. }}
\label{fig:positions}
\end{figure*}

In the integrated spectrum, we identify two emission lines, which we fit as Gaussians (Fig. \ref{fig:mosaic}, right). The two Gaussian emission peaks are at ${V_{\rm{sys}}} = -623 \pm 30\ \rm{ km \ s^{-1}}$ with velocity $\rm{FWHM} = 179 \pm 71\ \rm{ km \ s^{-1}}$, and ${V_{\rm{sys}}} = -116 \pm 40 \ \rm{ km \ s^{-1}}$ with velocity $\rm{FWHM} = 346 \pm 87\ \rm{ km \ s^{-1}}$. We show these fits as the blue and the red Gaussians, respectively,  in Fig. \ref{fig:mosaic}. The zero point ${V_{\rm{sys}}} = 0 \ \rm{ km \ s^{-1}}$ in the spectrum corresponds to a redshift $z = 1.44$, the mean confirmed cluster member redshift from \citet{2018ApJ...859...38N}, as explained in the observation section. In Appendix~\ref{peaks}, we identify the emission regions of the two CO(2-1) emission peaks by mapping the position of each component using the GILDAS software and find that the two peaks correspond to two separate regions, one south-east and the other south-west of the AGN. Hereafter we identify the two peaks as the eastern and western emission peaks.

In Fig. \ref{fig:positions}, we compare the emission regions that we find from this analysis to the position of the CARLA IRAC color-selected galaxies in our {\it HST}/F140W image. We find that the spatial extension that corresponds to the eastern and western emission peaks is south of central AGN.  Neither the eastern or western emission peaks correspond to the spatial position or to the spectroscopic redshift of the AGN (Fig. \ref{fig:mosaic}). The peaks  do not correspond to any optical ({\it HST}/WFC3) or infrared ({\it Spitzer}/IRAC) counterpart.   We remind the reader that the bright central source south of the AGN is a star and not a galaxy (see Sec.~\ref{carla}).

The other detections at the center of the NOEMA CO(2-1) line emission mosaic map (Fig. \ref{fig:continuum}, right) are at $ \rm{SNR} \leq 2  $. Their velocity peak is at the same spectral position as the western peak, and we again do not detect galaxies at their position in the {\it HST}/WFC3 or {\it Spitzer}/IRAC images. We conclude that those overdensities might be due to the side lobes, and neglect them.
We do not have detections at $> 3 \sigma$ at the edges of the mosaic, where the noise is higher.

\begin{figure}[h]
\centering
\includegraphics[width=\hsize]{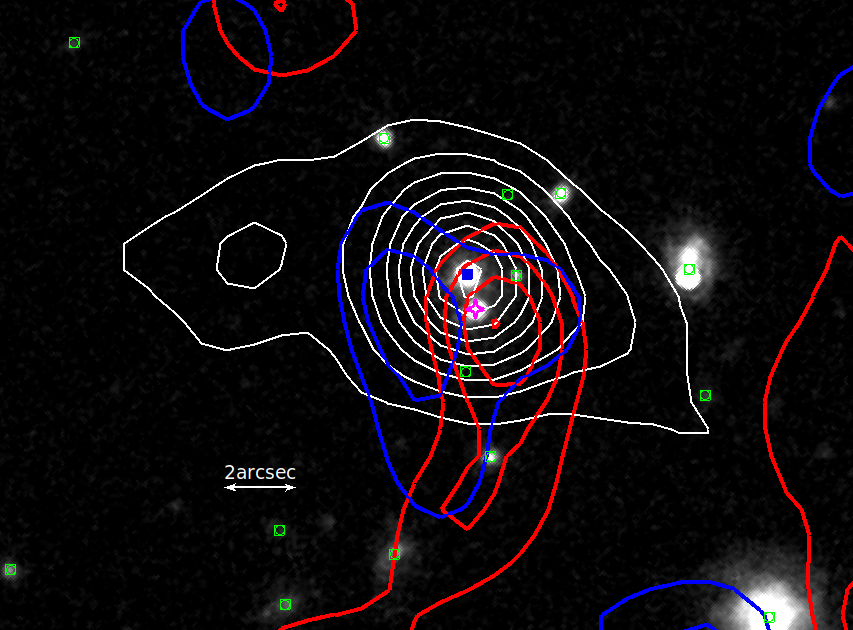}
\caption{\textit The {\it HST/WFC3 F140W image of the central region of the cluster, with the contours of CO(2-1) emission of the eastern (1-2~$\sigma$) and western (1-4~$\sigma$) emission peaks (shown as blue and red contours, respectively), and the continuum emission at 94.5 GHz (white contours; 3-24~$\sigma$).  The central blue square and the pink star with four spikes show the position of the AGN and of the star, respectively.  The green squares are the positions of IRAC color-selected galaxies in the cluster central region.  North is up and East is to the left.}}
\label{fig:cont_co}
\end{figure}

\paragraph{Flux and Luminosity.} \label{fluxlum}

From the Gaussian fit, the velocity integrated flux for the eastern and western emission peaks are $S_{\rm{CO(2-1)}} \Delta v = 0.6 \pm 0.2 \rm{ \ Jy \ km \ s^{-1}}$ (SNR$\sim$3; here and hereafter the SNR is calculated as the signal divided by its uncertainty, before approximating to one significant figure), and $S_{\rm{CO(2-1)}} \Delta v = 0.9 \pm 0.2 \rm{ \ Jy \ km \ s^{-1}}$ (SNR$\sim$4), respectively.  In the Gaussian fit, we leave all parameters free to vary.  The uncertainty on the measurements includes the uncertainties in the Gaussian fit and the noise in the region in which the fit is performed.  The total flux for the eastern emission peak, obtained by integrating over the velocity range $-1075 \  \rm{km \ s^{-1}} < v < -475 \ km \ s^{-1}$, is $S_{\rm{CO(2-1)}} \Delta v = 0.88  \pm 0.16 \rm{ \ Jy \ km \ s^{-1}}$ (SNR$\sim$6; Fig.~\ref{fig:positions};  Fig.~\ref{fig:cent2}). The difference between this flux measurement and that obtained from the Gaussian fit is consistent with zero. The total flux for the western emission peak, obtained by integrating over the velocity range  $-375 \ \rm{km \ s^{-1}} < v < +125 \ km \ s^{-1}$, is $S_{\rm{CO(2-1)}} \Delta v = 0.90 \pm 0.15 \rm{ \ Jy \ km \ s^{-1}}$ (SNR$\sim$6; Fig.~\ref{fig:positions};  Fig.~\ref{fig:cent3}), and very similar to the value obtained from the Gaussian fit.  The integrated flux from the eastern and western emission peaks over the velocity range $-1075\  \rm{km \ s^{-1}} < v < 125  \ \rm{km \ s^{-1}}$  is  $S_{\rm{CO(2-1)}} \Delta v = 1.7 \pm 0.2 \rm{ \ Jy \ km \ s^{-1}}$ (SNR $\sim 10$). We do not measure any CO(2-1) line emission at the spectral position of the AGN, $z = 1.4427 \pm 0.0005$, which corresponds to a velocity of $v = 331.6 \ \rm{ km\ s^{-1}}$, shown as a vertical green line in Fig.~\ref{fig:mosaic}. Hereafter, we use the velocity integrated fluxes for both the eastern and western emission peaks, for which we have the higher SNR.
We calculate the CO(2-1) luminosity, using the following relation from Eq. (3) of  \cite{2005ARA&A..43..677S} :

\begin{equation}
L'_{\rm{CO(2-1)}} = 3.25 \times 10^7 \frac{S_{\rm{CO(2-1)}} \Delta v D_{L}^2} {{\nu_{\rm{rest}}^2} (1+z)}
\end{equation}%
where $L'_{\rm{CO(2-1)}}$ is the CO(2-1) line luminosity  in $ \rm{\ K\ km\ s^{-1} \ pc^2}$, $S_{\rm{CO(2-1)}} \Delta v$ is the CO(2-1) velocity integrated flux in $\rm{Jy \  km \ s^{-1}}$, $D_L = 10397.4 \ \rm{Mpc}$ is the AGN luminosity distance, $\nu_{\rm{rest}} = 230.538\ \rm{GHz}$ is the rest frequency of the CO(2-1) rotational transition, and $z = 1.4427 \pm 0.0005$ is the AGN redshift (see Sec.~\ref{carla}). 
We find $L'_{\rm{CO(2-1)}} = 2.4 \pm 0.4 \times 10^{10} \rm{ \ K \ km \ s^{-1}  pc^2}$ and $L'_{\rm{CO(2-1)}} = 2.4 \pm 0.4 \times 10^{10} \rm{ \ K \ km \ s^{-1}  pc^2}$, for the eastern and western emission peaks, respectively.

\paragraph{Molecular gas mass.} \label{CO}

In order to estimate the molecular gas mass, we use the mass-to-luminosity relation:
\begin{equation}
M_{\rm{gas}} = \alpha_{\rm{CO}} \frac{L'_{\rm{CO(2-1)}}} {r_{21}} 
\end{equation}%
where $M_{\rm{gas}}$ is the molecular gas mass, {$\alpha_{\rm{CO}}$} is the CO-to-H$_2$ conversion factor (e.g., see review by \citealp{2013ARA&A..51..207B}), $r_{21}$ is the $L'_{\rm{CO(2-1)}}/L'_{\rm{CO(1-0)}}$ luminosity ratio, and $L'_{\rm{CO(2-1)}}$ and $L'_{\rm{CO(1-0)}}$ are the luminosities of the CO(2-1) and CO(1-0) emission lines, respectively. 

We assume thermalized, optically thick CO emission for which the CO luminosities are independent of the rotational transitions, thus, $L'_{\rm{CO(2-1)}} = L'_{\rm{CO(1-0)}} \equiv L'_{\rm{CO}}$ and $r_{21} = 1$ (\citealp{2005ARA&A..43..677S}). This is a standard value used for the local galaxy M82 (\citealp{2005A&A...438..533W}) and for color-selected star-forming galaxies (CSGs, \citealp{2009ApJ...698L.178D}). However other works use different values of $r_{21}$, such as $r_{21} = 0.5$ for the Milky Way (\citealp{2005A&A...438..533W}). The reader should take into account these differences when comparing to other works in the literature (e.g., \citealp{2013A&A...558A..60C}; \citealp{2017ApJ...842L..21N}; \citealp{2017ApJ...849...27R};  \citealp{2018ApJ...856..118H};  \citealp{2018arXiv180509789C};  \citealp{2018arXiv180601826C}).

The  CO-to-H$_2$ conversion factor $\alpha_{\rm{CO}}$ is a large uncertainty in this calculation. Its value is not universal and depends on galaxy type, metallicity and CO gas excitation temperature and density (\citealp{2013ARA&A..51..207B}; \citealp{2013ARA&A..51..105C}; \citealp{2018A&ARv..26....5C}). 
 For different kinds of galaxies and environments, its average range of values is  $0.8<\alpha_{\rm{CO}}<4.36 \ \ M_{\astrosun} \  {(\rm{ K \ km \ s^{-1} pc^2}})^{-1}$ \citep{2013ARA&A..51..207B}. 
 Since neither the eastern nor the western emission peaks are associated with galaxies detected in our optical and near-infrared images, they might be associated with extended emission around the AGN. In that case, we expect that the molecular gas might be more excited and with more chaotic motions, and this might lead to an expected value of $r_{21} > 1$, and to  $\alpha_{\rm{CO}} < 4.36 \ M_{\astrosun} \ (\rm{ K \ km \ s^{-1} pc^2})^{-1}$ (\citealp{2013ARA&A..51..207B}; \citealp{2013ARA&A..51..105C}; \citealp{2018ApJ...863..143C}).  For these reasons, we use the lower end of standard $\alpha_{\rm CO}$ values, and this will give us lower limits to the molecular gas mass. When using $\alpha_{\rm CO} = 0.8 \ M_{\astrosun} \ (\rm{ K \ km \ s^{-1} pc^2})^{-1}$, we obtain $M^{\rm{eastern}}_{\rm{gas}}= 1.9  \pm 0.3 \times 10^{10} M_{\astrosun}$ and $M^{\rm{western}}_{\rm{gas}} = 2.0 \pm 0.3 \times 10^{10} M_{\astrosun}$, for the eastern and western peak components, respectively. When using  the Galactic conversion factor  {$\alpha_{\rm{CO}} = 4.36 \ M_{\astrosun} \ {(\rm{ K \ km \ s^{-1} pc^2}})^{-1}$}, the molecular gas masses are $\sim 5$ times larger.
Summing the two components, the total molecular gas mass is $M^{\rm{tot}}_{\rm{gas}} = 3.9  \pm 0.4 \times 10^{10} M_{\astrosun}$ (SNR$\sim$8), and this mass is not spatially associated with galaxies detected in our optical or near-infrared images. Table \ref{tab:B} shows the integrated flux measurements, the CO luminosity and the molecular gas mass. 

\begin{table*}[h]
\center
\caption{Integrated flux measurements, the CO luminosity and the gas mass from the integrated CO(2-1) line emission.}
\begin{tabular}{c c c c c c c c c}
\hline\hline 
\\
Peak & ${S_{\rm{CO}} \Delta v} $ & $L'_{\rm{CO}} $  & $M_{\rm{gas}} $  \\
 & ($\rm{Jy \ km\ s^{-1}}$) & $(10^{10} \ \rm{Kpc^2km s^{-1}})$  & $(10^{10}M_{\astrosun})$  \\
\\
\hline Eastern &   $0.88 \pm 0.14$ & $2.4 \pm 0.4 $ & $1.9\pm 0.3$   \\
\hline Western &  $0.90 \pm 0.14 $ &$2.4 \pm 0.4 $ & $2.0 \pm 0.3$ \\\hline \\
\end{tabular} \\
\label{tab:B}
\tablefoot{Since we used the lower end of standard $\alpha_{\rm CO}$ values, we show lower limits to the molecular gas. }
\end{table*}

\subsection{MOLECULAR GAS CONTENT AND STAR FORMATION IN CLUSTER CORE MEMBERS}

In this section, we present star formation properties of spectroscopically confirmed cluster members that are in the region covered by the NOEMA observations.

\subsubsection{Upper limits on the molecular gas content of the cluster confirmed members} 

Besides the AGN, there are seven other spectroscopically confirmed CARLA J1103+3449 cluster members (\citealp{2018ApJ...859...38N}), of which six are within the NOEMA beam, and three have stellar mass estimates (Fig. \ref{fig:cluster_noirot}).  Our NOEMA observations do not show CO(2-1) emission with SNR $> 3$ at the positions of the spectroscopically confirmed members. However, we can use the $3\sigma$ values of the flux rms noise level around the position of each confirmed cluster member  to derive an upper limit on the velocity integrated flux $S_{\rm{CO(2-1)}} \Delta v = (3 \sigma_{\rm{rms}}) \Delta v$. As  $\Delta v$, we used an average  $\Delta v = 300 \ \rm{ km \ s^{-1}}$, following \citet{2017ApJS..233...22S}. Since the velocity resolution of our CO(2-1) map has an uncertainty of $\rm{\sigma_{\Delta v}} = 100 \ km \ s^{-1} $, the velocity range within $300 \pm \rm{3\sigma_{\Delta v}} \ \rm{ km \ s^{-1}}$ includes most of the published $\Delta v$ for star-forming cluster galaxies at these redshifts (e.g.,  \citealp{2017ApJ...842L..21N}; \citealp{0004-637X-842-1-55}; \citealp{2018arXiv180601826C}; \citealp{2018ApJ...856..118H}).   
 For the molecular gas measurement, we use the Galactic conversion factor  {$\alpha_{\rm{CO}} = 4.36 \ M_{\astrosun} \  {(\rm{ K \ km \ s^{-1} pc^2}})^{-1}$}, as typical for normal galaxies (\citealp{2013ARA&A..51..207B}; \citealp{2013ARA&A..51..105C}; \citealp{2018A&ARv..26....5C}). The estimated physical properties of the spectroscopically confirmed members are given in Table \ref{tab:E}.

\subsubsection{Star Formation Rates}

In this subsection, we calculate galaxy star formation rates using the H${\alpha}$ emission line flux from \citet{2018ApJ...859...38N}. We then combine them with our measurements of the molecular gas mass from the CO(2-1) line emission, and the galaxy stellar masses from Mei et al. (in preparation), in order to estimate the galaxy gas fraction, depletion time, star formation efficiency (SFE) and specific star formation rate (sSFR). 

\paragraph{Galaxy stellar masses and gas fractions.}
Mei et al. (in preparation) describe the details of our stellar mass measurements.
We measure our CARLA galaxy stellar masses  by calibrating our PSF-matched {\it Spitzer}/IRAC1 magnitudes (Amodeo et al., in preparation) with galaxy stellar masses from \cite{2015ApJ...801...97S} derived from the  \cite{2013ApJS..207...24G} multi-wavelength catalog in the Cosmic Assembly Near-infrared Deep Extragalactic Legacy Survey (CANDELS; PI: S. Faber, H. Ferguson; \citealp{2011ApJS..197...36K}; \citealp{2011ApJS..197...35G}) WIDE GOODS-S field.

The {\it Spitzer} IRAC1 magnitudes correspond to the rest-frame near-infrared in the redshift range of the CARLA sample, and we expect them not to be  biased by extinction.  We find a very good correlation between these magnitudes and the  \cite{2015ApJ...801...97S} mass measurements, with scatters of $\approx 0.12$~dex at the redshift of the cluster studied in this paper. Adding in quadrature the scatter of the relation and uncertainties from \cite{2015ApJ...801...97S}, we obtain mass uncertainties in the range $\sim 0.4-0.5$~dex, and  $\approx 0.2-0.3$~dex for masses larger than log$_{10} (\frac{M}{M_{\astrosun}}) > 10.5$. 

Table \ref{tab:E} shows the stellar masses of the cluster spectroscopically confirmed members.
The masses derived from this calibration are on average $\approx0.5$~dex smaller that those derived from stellar populations models by \cite{2018ApJ...859...38N}, and the difference is larger at fainter magnitudes.
This difference in mass estimates does not significantly change results from  \cite{2018ApJ...859...38N}, in particular the conclusions from the SFR vs. stellar mass analysis (Fig. 7 in \citealp{2018ApJ...859...38N}).
From our molecular gas mass upper limits, combined with our stellar masses, we compute the gas-to-stellar mass ratio as $M_{\rm{gas}}/M_{*}$ and the molecular gas fraction as $f_{\rm{gas}} = M_{\rm{gas}}/(M_{\rm{gas}} + {M_*})$. The results are shown in Table \ref{tab:E}.

\begin{table*}[h]
\center
\caption{Velocity integrated CO(2-1) flux,  luminosity, molecular gas mass, stellar mass, molecular gas-to-stellar mass ratio and molecular gas fraction for the CARLA J1103+3449 cluster confirmed members.}
\begin{tabular}{c c c c c c c c c c c}
\hline\hline 
\\
id & ${S_{\rm{CO}} \Delta v} $ &  $L_{\rm{CO}} \ $  & $M_{\rm{gas}} $ & $M_{\rm{*}} $  & $M_{\rm{gas}}/M_{*}$ &  $f_{\rm{gas}}$ \\
 & $(\rm{Jy \ km\ s^{-1}})$ &  $(10^{9} \ \rm{Kpc^2km s^{-1}})$  & $(10^{10}M_{\astrosun})$ & $(10^{10}M_{\astrosun})$  &  &  $(\%)$ \\
\\
\hline AGN/491 & $ < 0.3 $  &  $ <7 $ & $ < 3 $ & $5 \pm 2$ & $< 0.6 $ & $< 40$ \\
\hline 490 & $ < 0.3 $  & $ < 7 $ & $< 3 $ & ... & ... & ... \\
\hline 320 &  $ < 0.4 $ & $< 10 $ & $< 4$  & $0.7 \pm 0.7$ & $<6$ & $ <85$ \\
\hline 283  &$ < 0.5$ & $< 12$ & $< 5 $ &...&...&...\\
\hline 279  &$ < 0.5$ & $< 12$ & $< 5 $ &...&...&...\\
\hline 199  &  $ < 0.6$ & $< 17 $ & $< 7$ & $4 \pm 2$ & $< 2$ & $< 66 $ \\
\hline 183  &  $ < 0.6 $ & $< 17 $ & $< 7$ & $0.3 \pm 0.3$ & $<23$ & $<96$ \\
\hline 
\end{tabular}
\label{tab:E}
\tablefoot{The identification numbers in the column "id" are the same as in the catalog published by \cite{2018ApJ...859...38N}. The other columns show the velocity integrated CO(2-1) flux,  luminosity, molecular gas mass, stellar mass, molecular gas-to-stellar mass ratio and molecular gas fraction of the  CARLA J1103+3449 spectroscopically confirmed cluster members in the cluster core. The spectroscopically confirmed members were not detected with NOEMA and we report their $3\sigma_{\rm{rms}}$ upper limits.}
\end{table*}

\begin{table*}[h]
\center
\caption{ $\rm{SFR_{H{\alpha}}^{\rm{N18}}} $, attenuation, metallicity, SFR, sSFR, depletion time, and SFE for the CARLA J1103+3449 cluster confirmed members. }
\begin{tabular}{c c c c c c c c c c c }
\hline\hline
\\
 id & $\rm{SFR_{H{\alpha}}^{\rm{N18}}} $ &$A_{H\alpha}$ &12 + log(O/H) & $\rm{SFR_{H{\alpha}}} $ & sSFR  & $\rm{\tau_{dep}} $ &  $\rm{SFE} $ \\\\
 & $(M_{\astrosun}\rm{yr^{-1}})$ & && $(M_{\astrosun}\rm{yr^{-1}})$ & ($\rm{Gyr^{-1}}$)&$(\rm{Gyr})$& $(\rm{Gyr^{-1}})$\\
 \\
\hline AGN/491 $100\%$ &<140 & 1.4 & 8.6 & $140 \pm 50$ & $3 \pm 2$ &  $< 0.2$ & $>5$\\
\hline AGN/491 $80\%$ &...& 1.4 & 8.6 & $110\pm 40$ & $2 \pm 1$ & $< 0.3$ & $>4$\\
\hline AGN/491 $60\%$ & ...& 1.4 & 8.6 & $80 \pm 30$ & $2 \pm 1$ & $<0.4$ & $>3$\\
\hline AGN/491 $40\%$& ... &1.5 & 8.6 & $50 \pm 20$ & $1.1 \pm 0.7$ & $<0.6$ & $>2$\\
\hline AGN/491 $20\%$& ... &1.5 & 8.7& $30 \pm 8$ & $0.5 \pm 0.3$ & $<1$ & $>0.8$\\
\hline 490$^{(+)}$ &25 $\pm$ 5&...&...&...&...&$<1$ & $ > 0.8$ \\
\hline 320 &10 $\pm$ 5 & 0.8 & 8.6 & $6 \pm 3 $ & $1 \pm 1$ &$ <7$ & $> 0.1$\\
\hline  283$^{(+)}$ &16 $\pm$ 7 &...&... &...& ... &$< 3$ & $> 0.3$\\
\hline 279$^{(+)}$&11 $\pm$ 6&... &... .&...& ... & $< 5$ & $> 0.2$\\
\hline 199 & 11 $\pm$ 4& 1.4 & 8.7 & $9 \pm 4 $ & $0.2 \pm 0.1$ & $< 8$ & $> 0.1$\\
\hline 183 & 13 $\pm$ 5&0.5 & 8.4 & $6 \pm 4 $ & $2 \pm 3$ &  $< 12$ & $> 0.1$\\
\hline 93 &10 $\pm$ 2 &... &...&...&... &...&... \\
\hline
\end{tabular}
\tablefoot{The identification numbers in the column "id" are the same as in the catalog published by \cite{2018ApJ...859...38N}. The column $\rm{SFR_{H{\alpha}}^{\rm{N18}}} $ shows the SFR calculated in \cite{2018ApJ...859...38N}. The other columns show our measurements of  attenuation, metallicity, SFR, sSFR, depletion time, and SFE for the CARLA J1103+3449 cluster confirmed members.  In the case of the AGN (id 491), since we cannot separate the AGN and stellar contributions, we vary the stellar contribution to the total H${\alpha}$+[NII] emission flux in the range  $20 - 100\%$. The symbol (+) shows the cluster members for which we use the values of SFR reported in \cite{2018ApJ...859...38N} to estimate depletion time and SFE. }
\label{tab:F}
\end{table*}

\paragraph{Star formation rates, specific star formation rates, depletion times and star formation efficiencies.} \label{sec:sfr}

We re-compute $\rm{SFR_{H{\alpha}}}$, using the $\rm{{H{\alpha}}}$ line fluxes from \cite{2018ApJ...859...38N} and our stellar masses from Mei et al. (in preparation).
The Kennicut law \citep{1998ARA&A..36..189K} shows a direct proportionality between $\rm{SFR}$ and H${\alpha}$ flux:
\begin{equation}
\rm{SFR}_{\rm{H{\alpha}}} [M_{\astrosun} \rm{yr^{-1}}] =  5 \times 10^{-42} L_{\rm{H{\alpha}}} \times 10^{0.4 \times A_{\rm{H{\alpha}}}}
\end{equation}
where SFR$_{\rm{H{\alpha}}}$ is the estimated SFR corrected for the contribution from the [NII] line. $A_{\rm{H{\alpha}}}$ is the dust attenuation. 

We estimate $A_{\rm{H{\alpha}}}$ using the \citet{2010MNRAS.409..421G} empirical law (which used the  \citealp{2000ApJ...533..682C} extinction law):

\begin{equation}
A_{\rm{H{\alpha}}} = 0.91 + 0.77M + 0.11M^2 -0.09M^3
\end{equation}%

\noindent where $M = log_{10}{\frac{M_*}{10^{10} M_{\astrosun}}}$  and $M_*$ is the stellar mass.

\noindent $L_{\rm{H{\alpha}}}$ is the ${\rm{H{\alpha}}}$ luminosity in $\rm{erg \ s^{-1}}$, and it is calculated from $F_{\rm{H{\alpha}}}$, the H${\alpha}$ flux given in $\rm{erg \ \ cm^{-2} \ s^{-1}}$, which is computed as:

\begin{equation}
\rm{F_{H{\alpha}} = F_{H{\alpha} + [NII] \lambda 6548, 6584}\frac{1}{1 + \frac{F_{[NII] \lambda 6548, 6584}}{F_{H{\alpha}}}}}
\end{equation}%

\noindent $F_{\rm{H{\alpha} + [NII] \lambda 6548, 6584}}$ is the total observed H${\alpha}$ flux plus the [NII]$\lambda 6548, 6584$ flux. In fact, the WFC3/G141 grism resolution does not permit us to deblend the three lines \citep{2018ApJ...859...38N}.

To measure $ \rm{\frac{F_{[NII] \lambda 6548, 6584}}{F_{H{\alpha}}}}$, we use the relation between this ratio and metallicity, and the fundamental relation between stellar mass, SFR and metallicity. Following \citet{2020MNRAS.491..944C} (equations (2) and (5) and Table 6), we calculate the metallicity $\rm{12 + log(O/H)}$, expressed as a function of stellar mass and SFR:
\begin{equation}
\begin{aligned}
\rm{12 + log(O/H)}=  Z_0 - \gamma/\beta \times log \left[1 + \left(\frac{M_*}{M_0(\rm{SFR})}\right)^{-\beta}\right]
\end{aligned}
\end{equation}%
where $Z_0 = 8.779 \pm 0.005$, $log(M_0(\rm{SFR}))=m_0+m_1 \times log(\rm{SFR})$, $m_0 = 10.11\pm 0.03$, $m_1 = 0.56\pm0.01$, $\gamma = 0.31\pm0.01$, and $\beta = 2.1\pm0.4$.

\citet{2020MNRAS.491..944C} also provide a new calibration for the relation between metallicity and $\rm{\frac{F_{[NII]_{\lambda6584}}}{F_{H{\alpha}}}}$ :

\begin{equation}
log\left(\rm{\frac{F_{[NII]_{\lambda6584}}}{F_{H{\alpha}}}}\right) = \sum_{n=1}^{4} c_n x^n
\end{equation}%

\noindent where x ={\rm 12 + log(O/H) - 8.69},  $c_0 = -0.489$, $c_1 = 1.513$, $c_2 = -2.554$, $c_3 = -5.293$, and $c_4 = -2.867$.
Assuming a constant ratio ${\rm F_{[NII]_{\lambda6584}}}$:${\rm F_{[NII]_{\lambda6548}}}$ of 3:1 \citep{2006agna.book.....O}, we derive $ \rm{\frac{F_{[NII] \lambda 6548, 6584}}{F_{H{\alpha}}}}$. 

 Since to calculate SFR in Eq. 3 we need to know $ \rm{\frac{F_{[NII] \lambda 6548, 6584}}{F_{H{\alpha}}}}$, and to measure $ \rm{\frac{F_{[NII] \lambda 6548, 6584}}{F_{H{\alpha}}}}$ in Eq. 7 we need to know the SFR, we follow  \citet{2013ApJ...779..137Z} and start with an initial value of $ \rm{\frac{F_{[NII] \lambda 6548, 6584}}{F_{H{\alpha}}}}=0.2$ and iterate on Eq. 3-7 until convergence.

Our results are shown in Table~\ref{tab:F}.
For the AGN, we cannot separate the stellar contribution to the H${\alpha}$+[NII] line emission from the AGN contribution  \citep{2016A&ARv..24...10T}. 
Since we know that the hosts of powerful AGNs present young stellar populations (e.g., \citealp{2006NewAR..50..677H}),  we consider that the stellar contribution to the total H${\alpha}$+[NII] emission varies in the range of $20 - 100\%$.
In Table~\ref{tab:F}, we compare our SFR measurements with those from \cite{2018ApJ...859...38N}, and we are consistent within 1.5-2$\sigma$.
For some cluster galaxies we could not measure stellar masses, because they are not detected in the IRAC images, and we cannot re-compute the SFR. Hereafter, we will use our SFR measurements when we could derive them, and  \cite{2018ApJ...859...38N} SFR measurements for the other galaxies.
Combining our measured SFR with the stellar masses from Mei et al., we compute the specific star formation rate $\rm{sSFR} = \frac{SFR_{H{\alpha}}}{M_{*}}$, the depletion time $\rm{\tau_{dep}}= \frac{M_{\rm{gas}}}{\rm{SFR_{H{\alpha}}}}$, and the star formation efficiency  $\rm{SFE} = \frac{\rm{SFR_{H{\alpha}}}}{M_{\rm{gas}}}$.
Our results are shown in Table~\ref{tab:F}.

\section{Discussion} \label{Discuss}

\subsection{ORIGIN OF THE MOLECULAR GAS IN THE CLUSTER CORE}

In the core of the cluster, we observe two CO(2-1) emission peaks that correspond to a region spatially offset from the center of the AGN continuum emission and that does not correspond to any galaxy detected in our {\it HST} or {\it Spitzer} images. Radio observations of the CARLA J1103+3449 cluster from the work of \cite{1999MNRAS.303..616B} reveal two radio lobes, which are roughly in the same directions (east and west) as our CO(2-1) molecular gas components and our data extended continuum emission (Fig.~\ref{fig:contem},  \ref{fig:positions}, \ref{fig:cont_co}). The two radio lobes are asymmetrical, the western being more compact, while the eastern is more diffuse. The asymmetry of their widths may be due to their expansion in an ICM (Intracluster Medium) with a density gradient, in which the more diffuse lobe, the western lobe, would be expanding in a less dense environment (e.g., \citealp{2020PASA...37...13S}).
Both CO(2-1) emission peaks are blue-shifted compared to the NOEMA observation central velocity of $0 \ \rm{km \ s^{-1}}$ (which corresponds to the cluster redshift  of $z = 1.44$) and to the AGN redshift, and their spatial position is close to, but south of the AGN and the radio lobes. 

\subsubsection{Undetected galaxies}

To exclude the hypothesis that the two CO(2-1) line emission components might originate from two or more galaxies that are not detected at the detection limit of our optical or NIR images, we measure their hypothetical properties by making reasonable assumptions. 
Since our {\it HST}/WFC3 F140W images have a depth similar to the CANDELS WIDE survey, we use the CANDELS/WIDE survey mass limit  $M_{\rm{*}} < 5\times 10^{9}M_{\astrosun}$ (\citealp{2011ApJS..197...35G}; similar to our {\it Spitzer} mass limit) as an upper limit to the stellar mass of each of these two hypothetical galaxies.

Assuming the $3\sigma$ ${\rm H}{\alpha}$ emission line flux limit of ${\rm F_{H\alpha} = 2.1 \times 10^{-17} \ erg \ cm^{-2} \ s^{-1}}$ for the {\it HST}/WFC3 G141 grism spectra (\citealp{2016ApJS..225...27M}; \citealp{2018ApJ...859...38N}) as an upper limit of the undetected H$\alpha$ flux, and using the upper limit of the stellar mass, we estimate an upper limit to the SFR of the two emission peak components as $\rm{SFR_{H{\alpha}} < 2}\ M_{\astrosun} \ \rm{ yr^{-1}}$, using the same system of equations in Sec.~\ref{sec:sfr} and the cluster redshift\footnote{We assume that these hypothetical galaxies are at the cluster redshift because it would be very improbable to have two galaxies at another redshift so close to the cluster center and with spectral peaks so close to the cluster redshift.}. 
We then estimate lower limits to the molecular gas-to-stellar mass ratios and gas fractions that correspond to the eastern and western peak emission.  We obtain molecular gas masses  of $M^{\rm{blue}}_{\rm{gas}} = 10 \pm 3 \times 10^{10} M_{\astrosun}$, $M^{\rm{red}}_{\rm{gas}} = 11 \pm 3 \times 10^{10} M_{\astrosun}$.  We use the Galactic conversion factor because it is very improbable that these hypothetical galaxies are star-burst galaxies since they are not detected in our {\it HST}/WFC3 G141 grism observations. They could be only if the attenuation has an anomalously high values, i.e. $\rm A_{H{\alpha}} > 5$. Those gas masses lead to estimated gas fractions lower limits of 
$f_{\rm{gas}} \gtrsim 95\%$, and a lower limit on the depletion times of  $\rm{\tau_{dep}}  \gtrsim 55  \ \rm{Gyr}$. This is much longer than the depletion times observed for standard star-forming galaxies up to $z\sim4$, which are closer to $\sim 1-3$~Gyr (e.g., \citealp{2013ApJ...768...74T, 2018ApJ...853..179T}).
The probability that the line that we are observing is not CO(2-1) is very small, given that it is very close to the CO(2-1) emission expected at the cluster redshift. Some massive galaxies at $z>2$ can be detected in millimeter wavelength but not in the {\it HST} optical and near-infrared bands (e.g., \citealp{2018A&A...620A.152F}). However, these galaxies are rare (0.1 galaxy/arcmin$^2$), massive and usually detected with {\it Spitzer}/IRAC. Given the number densities of high redshift galaxies (e.g.,  \citealp{2018ApJ...852..107D}; \citealp{2018A&A...620A.152F}), having two galaxies of this kind so spatially close is possible but very improbable. These results mean that these two hypothetical galaxies would be unusually gas-rich, with low SFR (or anomalously high attenuation), high gas fractions and very long depletion times, independent of the conversion factor that we use. It is then very unlikely that our signal is due to undetected galaxies.

\subsubsection{Extended emission}

Excluding the hypothesis that the two CO(2-1) emission lines are due to undetected galaxies, they might trace molecular gas originating from an extended disk or torus, or emission components of molecular gas outflows or inflows associated with the AGN and its two radio jet lobes.
We find no evidence to support the hypothesis of CO emission from an extended (up to tens of $\rm{kpc}$) rotating disk or torus of molecular gas around the AGN.
In fact, the CO(2-1) eastern and western peak emissions are not spatially located at the AGN position, they are located south-east and south-west of the AGN and the radio jets. The total molecular gas mass in the southern structures around the AGN is $ \gtrapprox 60\%$ of the total molecular gas, from $M^{\rm{tot}}_{\rm{gas}} =  3.9 \pm 0.4 \times 10^{10} M_{\astrosun}$, and the upper limit on the AGN molecular gas mass ($<3 \times 10^{10} M_{\astrosun}$; Sect. \ref{CO}; Table~\ref{tab:E}). 

In the local Universe, both disk-dominated and filament-dominated central cluster galaxies were observed with ALMA  (\citealp{2019MNRAS.490.3025R}; \citealp{2019A&A...631A..22O}). In the first type, most of the molecular gas is concentrated in a disk around the central galaxy, while in the second type, the molecular gas is mostly ($>70\%$) in filaments around the central galaxy (the most known example being the Perseus cluster; \citealp{2006A&A...454..437S}). The filaments typically extend from a few kpc in length up to $10-20$ kpc, and the molecular gas emission is offset with respect to the central AGN by projected distances of a few kpc. For the central galaxy of the cluster A1795, some molecular gas clumps are associated with the lobes of the radio jets. In filament-dominated galaxies, the filaments trace radio bubbles, and are associated with both gas outflow and inflow. 

The molecular gas that we detect south of our central AGN also dominates the cluster central molecular gas reservoir and our observations are consistent with the filament-dominated local central galaxies. This suggests that the eastern and western emission peaks can be associated with gas outflow and inflow from the AGN.  
In fact, the AGN jets can drive a large amount of molecular gas, but this is not always expelled from the galaxy surroundings (e.g., \citealp{2015MNRAS.448L..30C}; \citealp{2015ApJ...811..108P}), and can be re-accreted. As a consequence,  the signal that we observe can be due to both inflows and outflows. When the amount of molecular gas outside the host galaxy is comparable to or higher than the host galaxy molecular gas reservoir, this also suggests that the gas has been cooled (e.g., \citealp{2004ApJ...612L..97K}; \citealp{2009MNRAS.395L..16N}; \citealp{2014MNRAS.438.2898E}; \citealp{2019MNRAS.490.3025R}).
For the CARLA J1103+3449 cluster, we expect a reservoir of hot intracluster medium (ICM) that surrounds the host AGN, with accretion at the center of the cluster potential well. The detection of cool molecular gas around a cluster central AGN can indicate cooling due to the interaction of the ICM and AGN jets or can be due to condensation of low entropy hot gas uplifted by the AGN jet away from the host galaxy (\citealp{2006A&A...454..437S};
\citealp{2008ApJ...672..252L}; \citealp{2012ApJ...746...94G};  \citealp{2014ApJ...785...44M}; 
\citealp{2014ApJ...784...78R};
\citealp{2014MNRAS.438.2898E};
\citealp{2015ApJ...811...73L};  \citealp{2015ApJ...811..108P}; 
\citealp{2017ApJ...837..149G}; 
\citealp{2017ApJ...845...80V}; \citealp{2018ApJ...865...13T};  \citealp{2019A&A...631A..22O}). With our observations, we cannot distinguish between these two scenarios.

\subsection{CLUSTER CONFIRMED MEMBER PHYSICAL PROPERTIES AND SCALING RELATIONS}

\begin{figure}[h]
\centering
\includegraphics[width=\hsize]{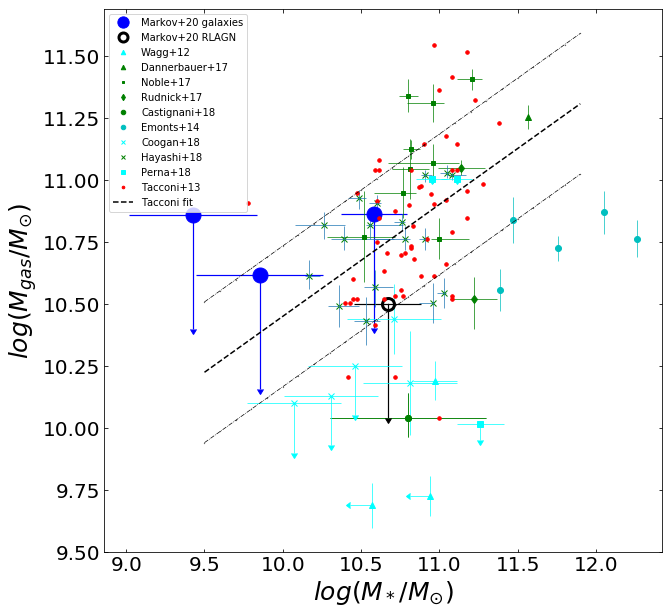}
\caption{\it Molecular gas mass vs. stellar mass relation for the AGN (empty black circles) and   for other CARLA J1103+3449 spectroscopically confirmed cluster members (blue filled circles) compared to other cluster (green and cyan) and field galaxies (red) from the literature.  The arrows show lower and upper limits. Green markers are results for cluster galaxies for which gas masses were estimated using the Galactic conversion factor. Cyan markers are estimations with different values of the conversion factor. Dashed and dashed-dotted lines represent the  \citet{2013ApJ...768...74T} relation for field star-forming galaxies and its 1$\sigma$ scatter.}
\label{fig:fgas}
\end{figure}

In this section, we compare our measured cluster confirmed member physical properties to both cluster and field galaxies at similar redshifts ($1 < z < 2.6$). Our  CO(2-1) luminosity, estimated upper limits to the velocity integrated CO(2-1) flux, and depletion times are similar to the literature for other cluster galaxies, AGN and spiral galaxies at $1<z<2.5$ (\citealp{2012ApJ...752...91W}; \citealp{2013A&A...558A..60C}; \citealp{2014MNRAS.438.2898E}; \citealp{2017ApJ...849...27R}; \citealp{2017ApJ...842L..21N}; \citealp{2018arXiv180601826C};  \citealp{2018ApJ...856..118H}; \citealp{2020arXiv200401786C}), and galaxies in the field in the same redshift range \citep{2013ApJ...768...74T}.
To compare our molecular gas upper limits to the literature, we have to take into account that we assume $L'_{\rm{CO(2-1)}}/L'_{\rm{CO(1-0)}} = 1$ (Sect. \ref{Results}; \citealp{2005ARA&A..43..677S}), and in several works a lower ratio of $\sim 20-50\%$ is assumed. For example, \cite{2017ApJ...842L..21N} use a ratio $L'_{\rm{CO(2-1)}}/L'_{\rm{CO(1-0)}} = 0.77$, \cite{2018ApJ...856..118H} use $L'_{\rm{CO(2-1)}}/L'_{\rm{CO(1-0)}} \sim 0.83$, and  \cite{2013ApJ...768...74T} assume $L'_{\rm{CO(3-2)}}/L'_{\rm{CO(1-0)}} \sim 0.5$. In Fig.~\ref{fig:fgas}-11, we show the original published values, without scaling. In fact, our results do not significantly change when using other values of $L'_{\rm{CO(2-1)}}/L'_{\rm{CO(1-0)}}$.
We also know that stellar mass estimations can differ up to a factor of $\sim 1.5-6$ ( $\sim 0.1-0.8$~dex) when using different techniques or different stellar population models (\citealp{2006ApJ...652...97V}; \citealp{2009ApJS..184..100L}; \citealp{2010MNRAS.407..830M}; \citealp{2011ApJ...732...12R}; \citealp{2012MNRAS.422.3285P}; \citealp{2018MNRAS.476.1532S}). 

In Fig.~\ref{fig:fgas}, we compare our molecular gas mass vs. stellar mass relation to other works. Our upper limits agree with the field molecular gas mass-to-stellar mass ratio from the PHIBBS survey \citep{2013ApJ...768...74T}.  This result also holds when considering the uncertainties in the $L'_{\rm{CO(2-1)}}/L'_{\rm{CO(1-0)}}$ conversion, conversion factor $\alpha_{\rm CO}$ and stellar masses. Our galaxies show upper limits that are higher than the molecular gas mass-to-stellar mass ratio in some of the other clusters. However, since they are only upper limits, we cannot make conclusions on environmental effects, apart from the fact that our cluster galaxies do not show evidence for larger gas reservoir than field galaxies with similar stellar mass.

\begin{figure}[h]
\centering
\includegraphics[width=\hsize]{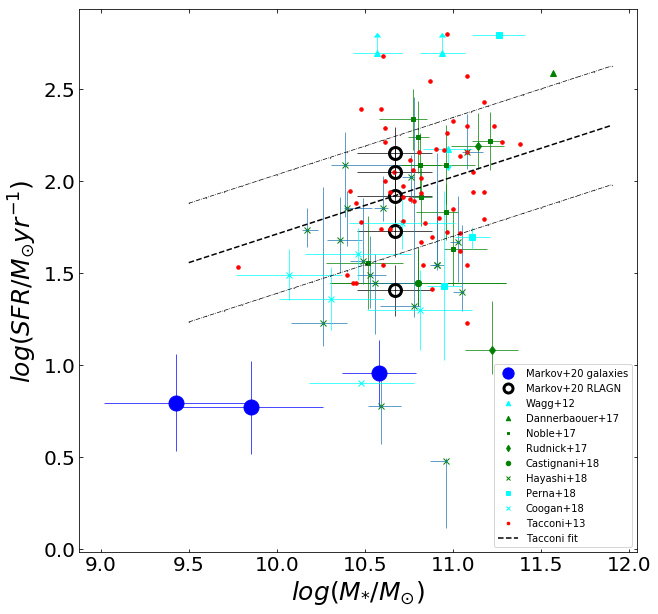}
\caption{\it SFR as a function of stellar mass. Symbols are the same as in Fig.~\ref{fig:fgas}. The AGN SFR is shown with different contributions of the H$\alpha$+[NII] stellar emission to the total flux.  We compare our results with those from other works. The dashed and the dashed-dotted lines represent the best fit with 1$\sigma$ scatter for the MS of field galaxies from \cite{2013ApJ...768...74T}. }
\label{fig:sSFR}
\end{figure}

\begin{figure}[h]
\centering
\includegraphics[width=\hsize]{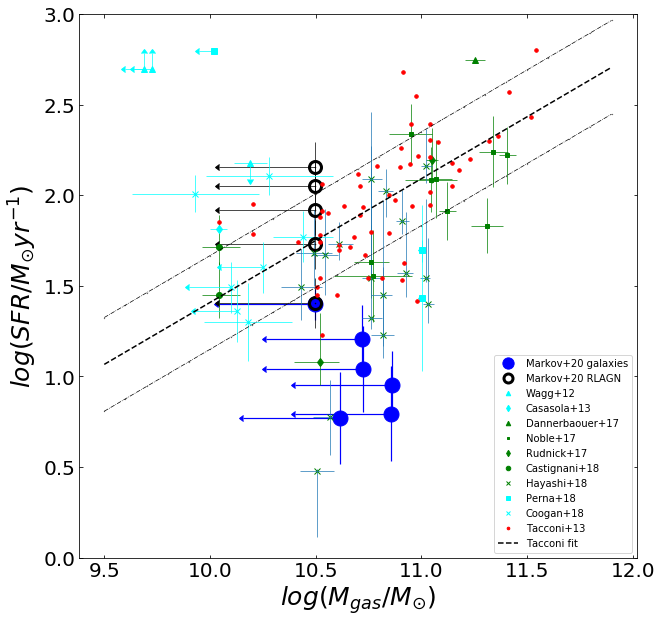}
\caption{ {\it SFR as a function of gas mass. Symbols are the same as in Fig.~\ref{fig:fgas}. The arrows show upper limits. Dashed and dashed-dotted lines represent the  \citet{2013ApJ...768...74T} relation for field star-forming galaxies and its 1$\sigma$ scatter.}}
\label{fig:SFE}
\end{figure}

In Fig.~\ref{fig:sSFR} and Fig.~\ref{fig:SFE}, we show SFR as a function of stellar and molecular gas mass, respectively.  Fig.~\ref{fig:sSFR} only shows results for galaxies for which we can measure the stellar mass. 
Compared to field galaxies, the SFR of the AGN host is within $\sim 1-1.5 \sigma$ of the main sequence (MS) from \cite{2013ApJ...768...74T}, and the SFR of the other spectroscopically confirmed cluster members is $\sim 2 \sigma$ lower than the MS (Fig.~\ref{fig:sSFR}), consistent with results from \citet{2018ApJ...859...38N}, which concluded that star-forming galaxies with stellar mass $>10^{10} M_{\astrosun}$ in the CARLA {\it HST} cluster sample have lower SFR than field galaxies with similar masses at the same redshifts.  In Fig.~\ref{fig:SFE}, the  AGN SFR (for all the H$\alpha$ stellar emission percentages considered in this paper) is also within $\sim 1 \sigma$ of field galaxies with gas masses similar to its molecular gas mass upper limit. This  shows that its SFR is typical of galaxies in the field with the same molecular gas reservoir.

The AGN SFR is comparable to the main sequence of star-forming field galaxies, and its star formation has not yet been quenched. This is consistent with the large molecular gas reservoir in the center of the cluster.
As a massive early-type central cluster galaxy, the AGN is predicted to have gone through at least one major merger, which might have triggered a starburst phase (\citealp{2006ApJS..163....1H, 2008ApJS..175..356H}; \citealp{2011ApJ...741...77S}; \citealp{2014ApJ...792...84Y}), which we do not observe as on-going in our data. During this phase, the galaxy is predicted to lie above the MS in the $\rm{SFR-M_*}$ and $\rm{SFR-M_{gas}}$ diagrams (Fig. \ref{fig:sSFR}, \ref{fig:SFE}), and the galaxy molecular gas is converted into stars. Afterwards, the remaining molecular gas content is expected to be consumed by the combination of star-formation and  feedback  (\citealp{2011ApJ...741...77S}; \citealp{2014ApJ...792...84Y}). Our AGN is observed on the star-forming galaxy MS (with a SFR of $\sim 30-140 M_{\astrosun} {\rm yr^{-1}}$), and we expect that it will evolve toward quenching when the molecular gas reservoir is depleted,  becoming a passive ETG similar to those observed in lower redshift cluster centers (\citealp{2001ApJ...557..150N}; \citealp{2008ApJS..175..356H}; \citealp{2011ApJ...741...77S}; \citealp{2014ApJ...792...84Y}).

\section{Summary} \label{Summary}

We report on observations of the central region of the galaxy cluster CARLA J1103+3449 at $z = 1.44$ with NOEMA, and measure the molecular gas content in the center of the cluster. We also obtain SFR, sSFR, and molecular gas mass,  SFE and gas depletion time upper limits for the spectroscopically confirmed cluster members.
\vskip 0.1cm
Our main results are:

\begin{itemize}

\item  At the rest frequency of $\nu_{\rm{rest}} =  230.5\ \rm{GHz}$ the dominant source of our NOEMA extended continuum emission is the non-thermal synchrotron radio emission from the AGN. We measure its flux at the AGN position and at the position of two radio jets. The central AGN in  CARLA J1103+3449 has been already observed at $4.71$ GHz and $8.21$ GHz by \cite{1999MNRAS.303..616B}, which found two asymmetrical radio lobes, one oriented towards the east and the other towards the west. We measure the continuum within a region the size of the NOEMA beam centered on the AGN and the two lobes, and obtain $S^{\rm{AGN}}_{\rm{cont}} = 4.6 \pm 0.2 \ \rm{mJy}$, $S^{\rm{east\_lobe}}_{\rm{cont}} = 1.1 \pm 0.2 \ \rm{mJy}$, and $S^{\rm{west\_lobe}}_{\rm{cont}} = 0.8 \pm 0.2 \ \rm{mJy}$. Combining our measurements with published results over the range 4.71~GHz-94.5~GHz, and assuming $\rm S_{synch} \propto \nu ^{-\alpha}$,  we obtain a flat spectral index $\alpha = 0.14 \pm 0.03$ for the AGN core emission, and a steeper index $\alpha = 1.43 \pm 0.04$ and $\alpha = 1.15 \pm 0.04$ at positions close to the western and eastern lobe, respectively. which is consistent with optically thicker  synchrotron emission.  The total spectral index is $\alpha = 0.92 \pm 0.02$ over the range 73.8~MHz-94.5~GHz.

\item   We detect two CO(2-1) emission line peaks with $\rm{SNR}\sim 6$, blue-shifted with respect to the AGN redshift.  One of the two detected emission peaks is situated at a projected distance of $\sim 17$~kpc  south-east of the AGN, and the second one is $\sim 14$~kpc south-west of the AGN. These regions are roughly aligned with the radio jets (east-west), and south of them.  These two emissions do not correspond to the position of any galaxy that we detect in our  optical or near-infrared images, and it is very unlikely that they are due to undetected galaxies (see the discussion section). 

\item We find a massive reservoir of cool molecular gas in the center of the cluster, distributed south of the AGN. From the  CO(2-1) total velocity integrated flux,  the total cluster core molecular gas mass is $M^{\rm{tot}}_{\rm{gas}} = 3.9\pm0.4\times 10^{10}M_{\astrosun}$.
The two CO(2-1) emission line peaks correspond to molecular gas masses of $M_{\rm{gas}} =  1.9\pm0.3 \times 10^{10}M_{\astrosun}$  for the eastern component, and of $M_{\rm{gas}} = 2.0\pm0.3  \times 10^{10}M_{\astrosun}$ for the western component. Considering the upper limit of $3 \times 10^{10}M_{\astrosun}$ on the AGN molecular mass (see below), the southern emission molecular gas mass is $\gtrapprox 60$\% of the cluster total central molecular mass reservoir. Our observations can be explained by gas inflows and outflows, either due to cluster gas accretion or, most probably, driven by the jets, as is observed in filament-dominated central galaxies in the local Universe. The gas might be cooled by the interaction of the ICM and AGN jets or can be due to condensation of low entropy hot gas uplifted by the AGN jet away from the host galaxy. 

\item The central AGN host is an ETG with a SFR of $\approx 30-140 \ M_{\astrosun} {\rm  yr^{-1}}$, depending on the assumed percentage of AGN contribution to its $\rm H{\alpha}$+[NII] flux (20\% to 100\%).
The upper limit on its gas reservoir is of $M_{\rm{gas}} < 3   \times 10^{10}M_{\astrosun}$. This means that the AGN molecular gas reservoir amounts to $ \lesssim 40\%$ of the total molecular gas reservoir in the center of the cluster.  The AGN host SFR lies on the MS of star-forming galaxies at similar redshift, and it has not yet been quenched. We expect that its star-formation will be also fed by the larger southern molecular gas reservoir.

\item  We measure SFR and sSFR, and estimate upper limits on the molecular gas masses, gas fractions, SFE and depletion times for the other spectroscopically confirmed cluster members. Our spectroscopically confirmed cluster member SFR is at $\sim 2 \sigma$ below the field star-forming MS (Fig.~\ref{fig:sSFR}), consistent with results from \citet{2018ApJ...859...38N}, which concluded that star-forming galaxies with stellar mass $>10^{10} M_{\astrosun}$ in the CARLA {\it HST} cluster sample have lower SFR than field galaxies at similar redshift, and of similar stellar mass. We find that the molecular gas mass upper limits are in the range of average values for field galaxies at similar redshifts and of similar stellar mass, and we cannot make conclusions on environmental effects apart from the fact that our cluster galaxies do not show evidence for larger gas reservoir than field galaxies with similar stellar mass.

\end{itemize}

\begin{acknowledgements}
We thank the PI of the Keck observations, Fiona Harrison, for providing these observations and Thomas Connor for assisting with the Keck observing.
We thank Philip Best and Katherine Inskip for useful discussion and their kind sharing of their radio, optical and infrared observations of the central radio sources and lobes. The data reduction and mapping and most of data analysis was done by using IRAM/GILDAS free software (http://www.iram.fr/IRAMFR/GILDAS/), and with the assistance of the IRAM support astronomers in Grenoble, Cynthya Herrera and Melanie Krips, which we warmly thank. We would like to thank the GILDAS support team for their help and guidance for the data analysis, S\'ebastien Bardeau and Vincent Pietu. V.M. would like to thank Anelise Audibert, Benoit Tabone, Valeria Olivares and Gianluca Castignani for their help with the data mapping and analysis. The work of DS was carried out at the Jet Propulsion Laboratory, California Institute of Technology, under a contract with NASA.  This work was supported by the CNES. \end{acknowledgements}

\bibliographystyle{aa}
\bibliography{biblio}

\begin{appendix} 

\section{The positions of the two CO(2-1) emission peaks} \label{peaks}

In order to kinematically resolve the two components of the CO(2-1) emission line, we perform an analysis of different ranges of channels (velocities or frequencies) in the CO(2-1) spectrum and create the corresponding CO(2-1) emission line intensity maps, averaged over the chosen range of velocities.  

\begin{figure*}[h]
\centering
\includegraphics[width=0.50\textwidth]{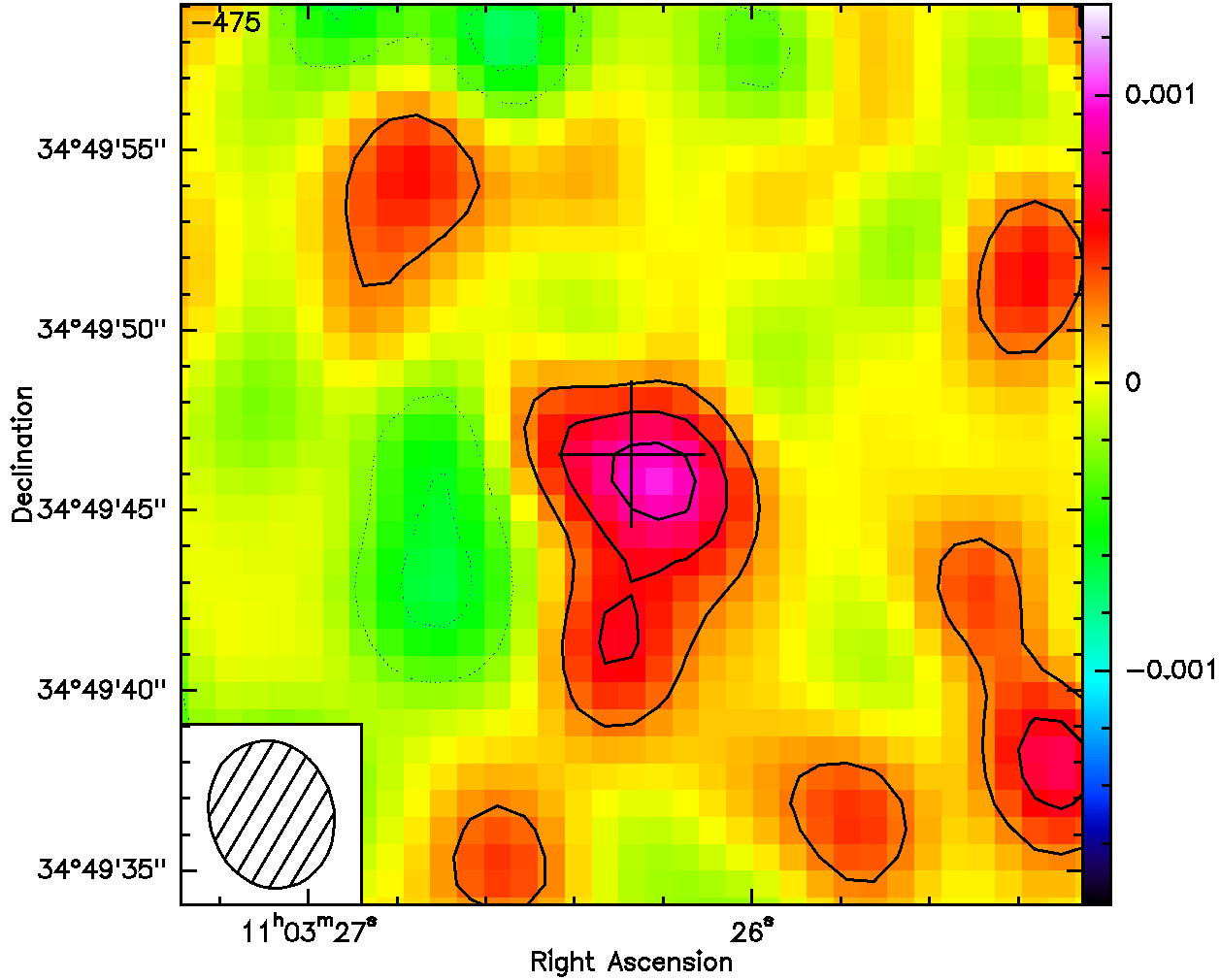}
\includegraphics[width=0.49\textwidth]{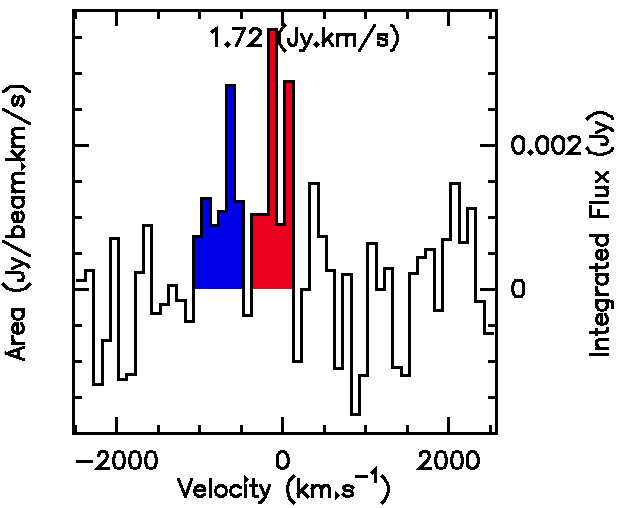}
\caption{\textit {Left: the CO(2-1) line emission intensity map of the central cluster region with NOEMA  created by selecting the channel ranges that include both the eastern and western emission peaks (on the right). The spectra are binned in channels of $100 \ \rm{km \ s^{-1}}$. The contour levels are $1\sigma$, $2\sigma$, and $3\sigma$.}}
\label{fig:cent1}
\end{figure*}

\begin{figure*}[h]
\centering
\includegraphics[width=0.50\textwidth]{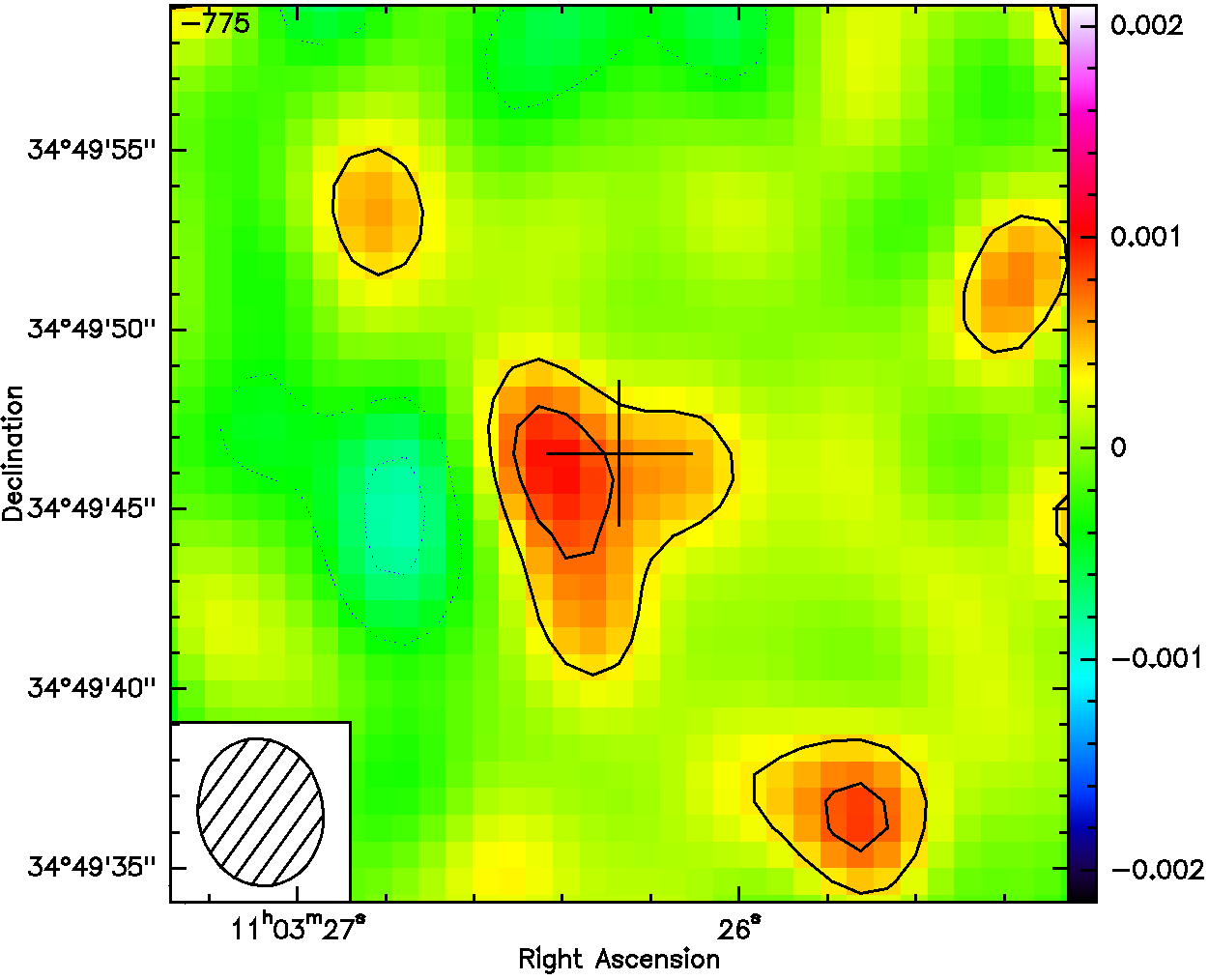}
\includegraphics[width=0.49\textwidth]{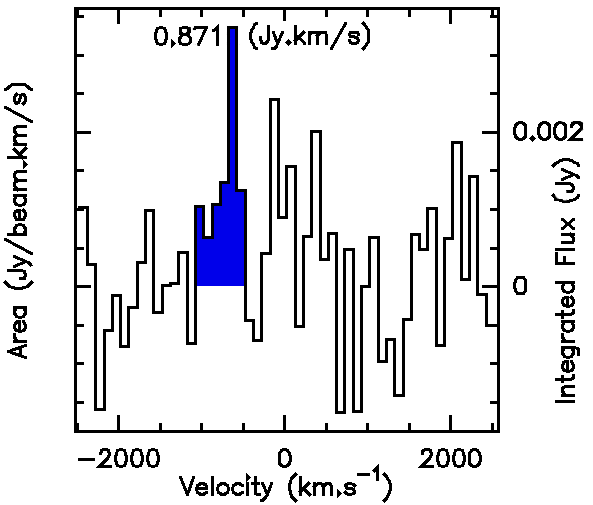}
\caption{\textit {Left: the CO(2-1) line emission intensity map of the central cluster region with NOEMA created by selecting the channel ranges that include only the eastern emission peak (on the right). The spectra are binned in channels of $100 \ \rm{km \ s^{-1}}$. The contour levels are $1\sigma$ and $2\sigma$.}}
\label{fig:cent2}
\end{figure*}

\begin{figure*}[h]
\centering
\includegraphics[width=0.50\textwidth]{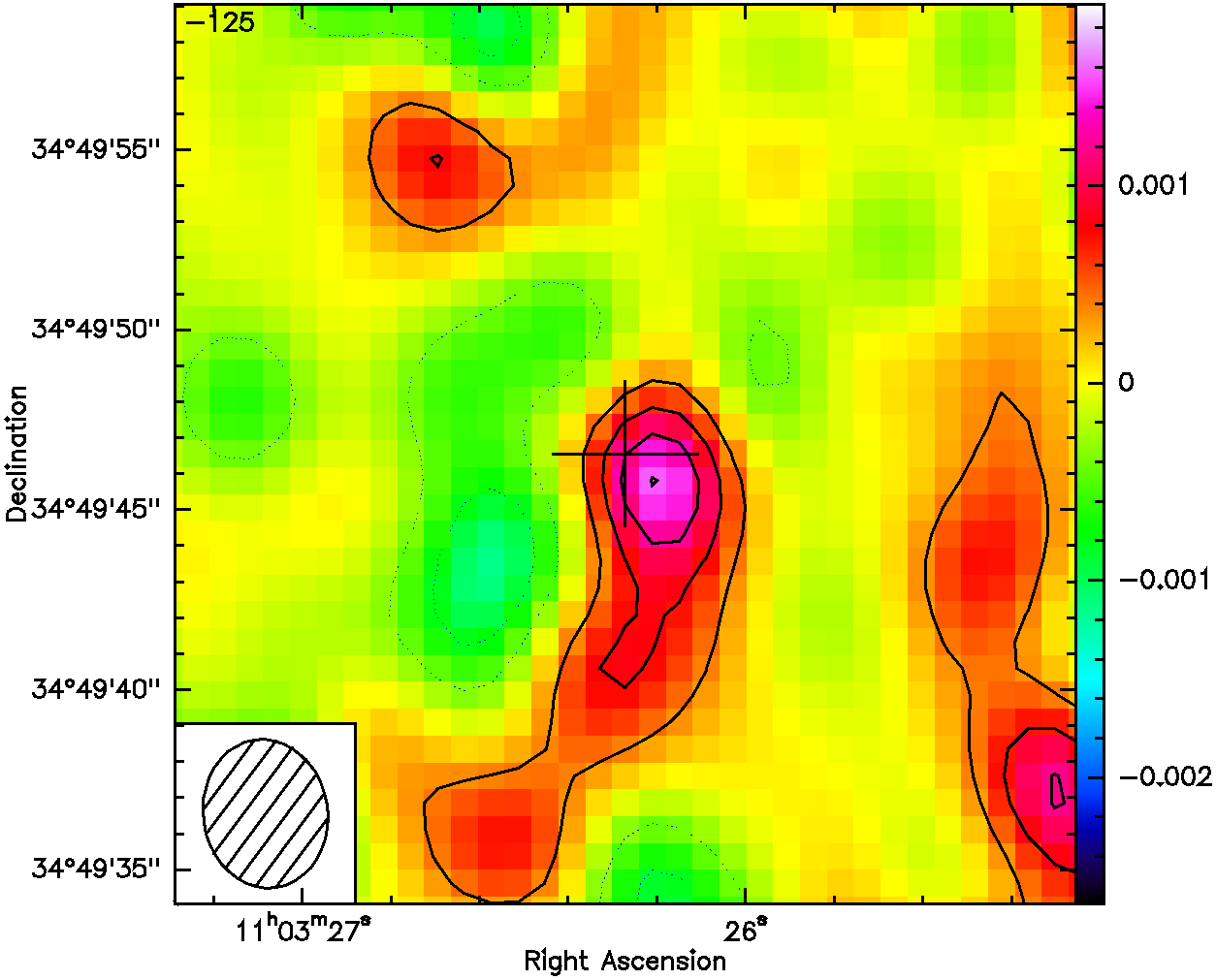}
\includegraphics[width=0.49\textwidth]{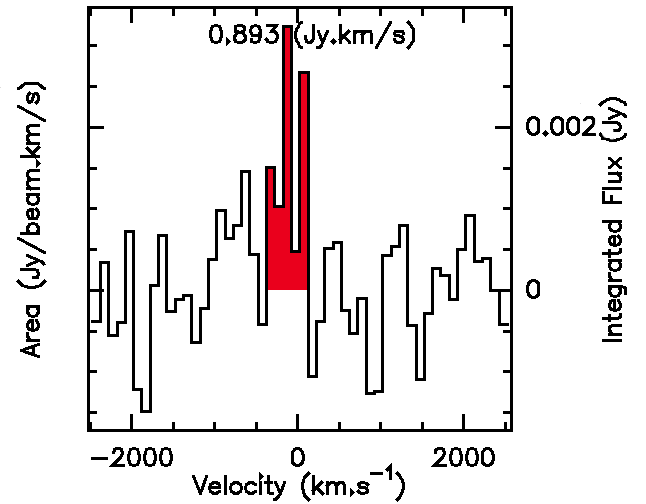}
\caption{\textit {Left: the CO(2-1) line emission intensity map of the central cluster region with NOEMA  created by selecting the channel ranges that include only the western emission peak (on the right). The spectra are binned in channels of $100 \ \rm{km \ s^{-1}}$. The contour levels are $1\sigma$, $2\sigma$, $3\sigma$ and $4\sigma$.}}
\label{fig:cent3}
\end{figure*}

In Fig.~\ref{fig:cent1}, we select a range of velocities so that we include both the eastern and western emission peaks ($-1075 \ \rm{km \ s^{-1}} < v < +125 \ km \ s^{-1}$. In Fig. \ref{fig:cent2}, we select only the range of velocities that correspond to the eastern peak ($-1075 \ \rm{km \ s^{-1}} < v < -475 \ km \ s^{-1}$).  In Fig. \ref{fig:cent3}, we select only the range of velocities that correspond to the western peak ($-375 \ \rm{km \ s^{-1}} < v < +125 \ km \ s^{-1}$). We can see that, although both the eastern and western emission peaks are superposed in some regions of the cluster core, we can kinematically separate them.

\clearpage

\end{appendix}

\end{document}